\documentclass[fleqn,usenatbib]{mnras}
\usepackage{newtxtext,newtxmath}
\usepackage[T1]{fontenc}
\DeclareRobustCommand{\VAN}[3]{#2}
\let\VANthebibliography\thebibliography
\def\thebibliography{\DeclareRobustCommand{\VAN}[3]{##3}\VANthebibliography}
\usepackage{graphicx}	% Including figure files
\usepackage{amsmath}	% Advanced maths commands

\usepackage{graphicx}	
\usepackage{amsmath}	

\title[A TDE candidate in High Redshift Quasar]{A central tidal disruption event candidate in high redshift quasar SDSS J000118.70+003314.0}

\author[Ying Gu et.al]{
Ying Gu,
Xue-Guang Zhang\thanks{E-mail: xgzhang@gxu.edu.cn},
Xing-Qian Chen, Xing Yang, 
and En-Wei Liang\thanks{E-mail: lew@gxu.edu.cn }
\\
Guangxi Key Laboratory for Relativistic Astrophysics, School of Physical Science and Technology,\\
Guangxi University, Nanning 530004, People's Republic of China\\
}

\date{Accepted XXX. Received YYY; in original form ZZZ}

\pubyear{\the\year{}}

\begin{document}

\label{firstpage}
\pagerange{\pageref{firstpage}--\pageref{lastpage}}
\maketitle

\begin{abstract}
We report a high-redshift ($z=1.404$) tidal disruption event (TDE) candidate in SDSS J000118.70+003314.0 (SDSS J0001), 
which is a  
quasar with apparent broad Mg~{\sc ii} emission line.
The long-term variability in its nine-year photometric $ugriz$-band light curves, obtained from the SDSS
Stripe82 and the PHOTOOBJALL databases,
can be described by the conventional 
TDE model. Our results suggest that the TDE is a main-sequence star with mass of $1.905_{-0.009}^{+0.023}{\rm M_\odot}$ tidally disrupted by a black hole 
(BH) with mass {$6.5_{-2.6}^{+3.5}\times10^7{\rm M_\odot}$}. The BH mass is about 7.5 times smaller than the virial BH mass derived from the broad Mg~{\sc ii} emission line, which can be 
explained by non-virial dynamic properties of broad emission lines from TDEs debris. Furthermore, we examine 
the probability that the event results from intrinsic variability of quasars, which is about $0.009\%$, through 
applications of the DRW/CAR process. Alternative explanations for the event are 
also discussed, such as the scenarios of dust obscurations, microlensing and accretion. Our results provide clues to support 
that TDEs could be detectable in broad line quasars as well as in quiescent galaxies, and to indicate the 
variability of some active galactic nuclei may be partly attributed to central TDEs.

\end{abstract}


%%%%%%%%%%%%%%%%%%%%%%%%%%%%%%%%%%%%%%%%%%%%%%%%%%

%%%%%%%%%%%%%%%%% BODY OF PAPER %%%%%%%%%%%%%%%%%%

\section{Introduction}
It is generally believed that massive galaxies host a central supermassive black hole (SMBH) 
\citep{2001AIPC..586..363K, 2013ARA&A..51..511K}. A tidal disruption event (TDE) 
occurs when a supermassive black hole rips apart a passing star \citep{1982ApJ...263..377N, 
1983A&A...121...97C}. TDEs are the best probes for studying the central SMBHs and corresponding accretion 
systems. They were theoretically predicted \citep{1982ApJ...263..377N,1983A&A...121...97C,1988Natur.333..523R} 
and were observationally confirmed (e.g. \citealt{1996A&A...309L..35B,1999A&A...349L..45K, 
2021ARA&A..59...21G}).

	Theoretically, the TDE phenomenon was firstly predicted by \citet{1988Natur.333..523R} and 
\citet{1989IAUS..136..543P}, who calculated the rate of fallback materials through numerical simulations 
and analytical methods by assuming a star completely destroyed by a SMBH. They have shown that approximately 
half of the star's mass can be captured and an accretion disk surrounding the SMBH is formed, leading to a 
bright flare event in the ultraviolet(UV)-optical and soft X-ray band  \citep{1988Natur.333..523R, 
1989IAUS..136..543P,1994ApJ...422..508K,2014ApJ...783...23G}. \citet{2001ApJ...549..467I,2004ApJ...615..855K} 
have proposed that TDEs should exhibit more complex behavior and partial disruption of stars is likely more 
common than the complete disruption. The numerical studies by \citet{2009ApJ...695..404R,2009MNRAS.392..332L} 
have given the time-dependent fallback rate for varying polytropic spheres with index $\gamma$. The simulations 
by \citet{2013ApJ...767...25G} show further effects of varying impact parameter $\beta$ on the time-dependent 
fallback rate in cases ranging from no mass loss to deeply penetrating encounters. Based on these theoretical 
results, the following two basic points can be obtained. First, a unique time-dependent
brightness decay pattern of $\sim{t^{-5/3}}$ is a common feature of optical TDEs at late times. Meanwhile, 
some studies have shown deviations from this pattern at early times after the time of peak brightness
such as \citet{2011MNRAS.410..359L, 2013ApJ...767...25G}.
Second, BH mass should be limited to be smaller than $\sim10^8{\rm M_\odot}$ (the Hills limits),
otherwise the star should be totally swallowed by the central SMBH before being tidally disrupted, unless the 
central SMBH spin is extremely fast \citep{2012PhRvD..85b4037K}. \citet{2014ApJ...783...23G} (the known TDEFIT 
code) have performed a hydrodynamical simulation to investigate the structures and the dynamical evolutions 
of the debris stream formed from the tidal disruption of a main-sequence star. Additionally, 
\citet{2019ApJ...872..151M} have added the TDE modular to the known MOSFIT code. 
More recent reviews on theoretical simulations of TDEs can be found in \cite{2019GReGr..51...30S}.

	Observationally, TDE candidates were first detected by the ROSAT satellite in the soft X-ray band 
\citep{1999A&A...343..775K,1999A&A...350L..31G,2000A&A...362L..25G}. Thereafter, similar X-ray TDE candidates
were also detected by other space X-ray missions, including the XMM-Newton observatory, Chandra
X-ray observatory, and Swift/X-Ray Telescope \citep{2002AJ....124.1308D,2007A&A...462L..49E,
2008A&A...489..543E, 2009A&A...495L...9C,2014ApJ...792L..29M,2015ApJ...811...43L,2017A&A...598A..29S,
2021SSRv..217...18S}. 
In addition to these X-ray TDEs, three on-axis jetted TDEs were discovered by the 
Neil Gehrels Swift Observatory: Swift J1644+57 \citep{2011Sci...333..203B, 
2011Natur.476..421B,2011Sci...333..199L}, Swift J2058.4+0516 \citep{2012ApJ...753...77C} and Swift J1112.28238 
\citep{2015MNRAS.452.4297B}. In addition, the most recent jetted TDE AT 2022cmc was discovered by an optical sky survey 
 \citep{2022Natur.612..430A,2023NatAs...7...88P,2024ApJ...965...39Y}.
A recent review on the fraction of TDEs that launch jets can be found in \cite{2020NewAR..8901538D}. Due to our main 
focus on optical TDEs, there is no further discussion on X-ray TDEs in the manuscript.

Thanks to the high-quality light curves provided by the public sky survey projects 
in optical bands, more and more optical TDEs have been reported through unique spectroscopic features and/or
through long-term variability properties of TDEs. Descriptions of one to two known TDE
candidates from each current sky survey are given as follows. Through SDSS (Sloan Digital Sky Survey) 
\citep{2023ApJS..267...44A} catalog of galaxies, several TDE candidates have 
been reported by \citet{2008ApJ...678L..13K, 2013ApJ...774...46Y} due to the discovered coronal lines. Meanwhile, through 
SDSS provided long-term photometric datasets in Stripe82 \citep{2008AJ....135..338F}, SDSS-TDE1 and 
SDSS-TDE2 have been reported in \citet{2011ApJ...741...73V}. Through CSS (Catalina Sky Survey) \citep{2009ApJ...696..870D},
one known optical TDE candidate has been reported in CSS100217:102913+404220 in \citet{2011ApJ...735..106D}. Through 
 OGLE (Optical Gravitational Lensing Experiment) \citep{2015AcA....65....1U, 2014AcA....64..197W}, 
the optical TDE OGLE17aaj has been reported in \citet{2019A&A...622L...2G, 2017MNRAS.465L.114W,
2020A&A...639A.100K}. Through ASAS-SN (All Sky Automated Survey for supernovae) \citep{2014ApJ...788...48S}, 
ASASSN-14ae and ASASSN-15lh have been reported in \citet{2014MNRAS.445.3263H, 2016NatAs...1E...2L, 2019MNRAS.488.4816W}. 
Through Pan-STARRS (Panoramic Survey Telescope \& Rapid 
Response System) medium-deep survey \citep{2007AAS...21114206C}, the known optical TDEs PS1-10jh and PS1-11af 
have been reported by \citet{2012Natur.485..217G, 2014ApJ...780...44C}. Through PTF (Palomar Transient Factory) \citep{2009PASP..121.1395L},
the optical TDE PTF09ge has been reported by \citet{2014ApJ...793...38A}. Additionally, through 
ATLAS (Asteroid Terrestrial-impact Last Alert System) \citep{2018PASP..130f4505T}, a faint and fast TDE in a quiescent 
Balmer strong Galaxy ATLAS18mlw was reported in \citet{2023MNRAS.519.2035H}. Through ZTF 
(Zwicky Transient Facility) \citep{2019PASP..131a8002B}, TDE AT2018zr/PS18kh have been reported by \citet{
2019ApJ...872..198V}.
The number of detected optical TDEs is increasing quickly. More than 100 optical TDE 
candidates have been reported in the literature (see the collected TDE candidates listed in \url{https://tde.space/}). 
The TDEs are rare and the rate is $\sim 10 ^{-4} - 10 ^{-5}$ year $^{-1}$ galaxy $^{-1}$ 
\citep{2016PhRvD..93h3005W, 2016MNRAS.455..859S}. Recently, several samples of TDE candidates have been identified, 
such as TDE candidates reported in \citet{2021MNRAS.508.3820S} from the eROSITA/SRG all-sky survey, and in 
\citet{2021ApJ...908....4V, 2023ApJ...942....9H} and \citet{2023ApJ...955L...6Y} from ZTF. The optical-UV 
properties of TDEs are discussed in \citet{2021ApJ...908....4V}. More recent reviews on observational properties 
of optical TDE candidates can be found in \citet{2021ARA&A..59...21G}.

%######################################################################################
 
	Besides the discussed photometric variability properties related to TDEs, spectroscopic 
properties of TDEs are discussed as follows. The optical spectra of TDEs can be roughly classified into 
three spectroscopic types in \citet{2021ApJ...908....4V},(1): TDE-H type, which has broad H${\alpha}$ and 
H${\beta}$ emission lines; (2): TDE-H+He, which is characterized by broad emission lines for H${\alpha}$ and H${\beta}$, 
a complex of emission lines around He~{\sc ii} $\lambda$4686, and often includes emission at N~{\sc iii} $\lambda$4640
and $\lambda$4100. Additionally, some cases exhibit emission at O~{\sc iii} $\lambda$3760; and (3): TDE-He, which has broad 
emission feature near He {\sc ii} $\lambda$4686\AA~ but no broad Balmer emission features. More recently, 
\citet{2023ApJ...942....9H} suggested the fourth spectroscopic class for TDEs, the TDE-featureless, which lacks clear 
features of the three classes above but can show host galaxy absorption lines.

  Besides optical spectroscopic features of optical TDE candidates, several TDEs have been well checked 
at UV-band: ASASSN-14li \citep{2016ApJ...818L..32C}, iPTF16fnl 
\citep{2018MNRAS.473.1130B}, PS16dtm \citep{2017ApJ...843..106B}, iPTF15af \citep{2017ApJ...846..150Y}, 
PS1-11af \citep{2014ApJ...780...44C}, PS18kh \citep{2019ApJ...880..120H}, AT2018zr \citep{2019ApJ...879..119H, 
2020ApJ...903...31H} and SDSS J014124+010306 \citep{2022MNRAS.516L..66Z}. Among the reported TDEs, besides the TDE candidate SDSS J014124+010306  reported by \citet{2022MNRAS.516L..66Z}, no one exhibits apparent broad Mg~{\sc ii} emission lines.
In addition, spectroscopic properties of both broad emission lines and continuum emissions related to central BH accreting process can be applied as apparent signs for broad line AGN (active galactic nucleus), however, those spectroscopic properties can also be expected from assumed central TDEs in host galaxies. Therefore, reporting a central TDE in a definitely normal broad line AGN with pre-existing AGN activity is still a challenge.

%**************************************************************************************************

Most reported optical TDE candidates are found in quiescent galaxies, partly because identifying TDEs occurring in AGNs is more difficult than in quiescent galaxies due to the intrinsic variability of AGNs, which often exhibit flare-like features (e.g. \citealt{1994ApJS...95....1E, 2017MNRAS.470.4112G}). Furthermore, transients in galaxies with known AGNs have been excluded from TDE search methods (e.g. \citealt{2023ApJ...942....9H,2021ApJ...908....4V,2023ApJ...955L...6Y}) to avoid being overwhelmed by spurious candidates. Several TDE candidates have been reported in AGNs. It has also been suggested that these TDEs may be related to the unique variability of AGNs. \citet{2015MNRAS.452...69M} reported that the changing-look 
AGN SDSS J015957.64+003310.5 \citep{2015ApJ...800..144L} could be related to a supersolar mass star disrupted 
by a $10^8$ $M_{\odot}$ SMBH, leading to the light curve decay being consistent with the 
$t^{-5/3}$ behavior. This feature is also found in a TDE candidate in  Changing-look AGN 1ES 1927+654 
\citep{2019ApJ...883...94T}. \citet{2015A&A...581A..17C} suggested that a flare in low-luminosity AGN IC 3599 
could be explained by a repeating partial TDE with a recurrence time of 9.5 years. \citet{2017ApJ...843..106B} 
found that the light curve of a luminous transient PS16dtm in a narrow-line Seyfert 1 galaxy can be fitted 
with the TDE model. \citet{2018MNRAS.475.1190Y} analyzed a long-duration TDE expected X-ray flare in the 
low-luminosity AGN NGC 7213. \citet{2018ApJ...857L..16S} proposed that the supersoft X-ray spectrum and the 
X-ray light curve of AGN GSN 069 can be well explained as a TDE. \citet{2018ApJ...859....8L} reported a 
candidate TDE in a N-rich quasar SDSS J120414.37+351800.5 via abundance ratio variability. Moreover, 
\citet{2020ApJ...903..116A} reported the first radio transient TDE CNSS J001947.3+003527, which is associated 
with the nucleus of a nearby S0 Seyfert galaxy at 77Mpc. \citet{2020ApJ...894...93L} reported a TDE candidate 
in SDSS J022700.7-042020.6 by analyzing its light curves.
\citet{2022A&A...660A.119Z} proposed that the outburst in an atypical narrow-line Seyfert 1 galaxy CSS 
J102913+404220 could be explained by a stellar TDE. In addition, \citet{2022MNRAS.516L..66Z} reported 
a TDE candidate located in the quasar SDSS J014124+010306 with an apparent broad Mg~{\sc ii} emission line.

Among the reported TDE candidates, almost all of them are at low redshift. However, 
utilizing high-redshift TDEs can provide further information on our understanding of the characteristics and 
evolution of circumstance of TDEs around central SMBHs in high redshift galaxies and/or AGNs. To date, only 
four TDEs are at redshift $z>1$, including Swift J2058.4+0516 at $z\sim1.185$ in \citet{2012ApJ...753...77C}, 
J120414.37+351800.5 at $z\sim2.359$ in \citet{2018ApJ...859....8L}, J014124+010306 at $z\sim1.060$ in 
\citet{2022MNRAS.516L..66Z}, AT2022cmc at $z\sim1.193$ in \citet{2022GCN.31602....1T}. Swift J2058.4+0516 has long-lasting X-ray emission and 
radio emission \citep{2012ApJ...753...77C}, SDSS J120414.37+351800.5 at $z\sim2.359$ is the highest redshift 
of TDE candidates, which is found in a N-rich quasars \citep{2018ApJ...859....8L}. AT 2022cmc was suggested as 
an optically bright and fast relativistic TDE \citep{2022Natur.612..430A, 2023NatAs...7...88P,2023ApJ...943L..18C,2023MNRAS.522.4028M,2023MNRAS.521..389R}. 

\begin{figure}
\centering
\includegraphics[width = 6cm,height=5cm]{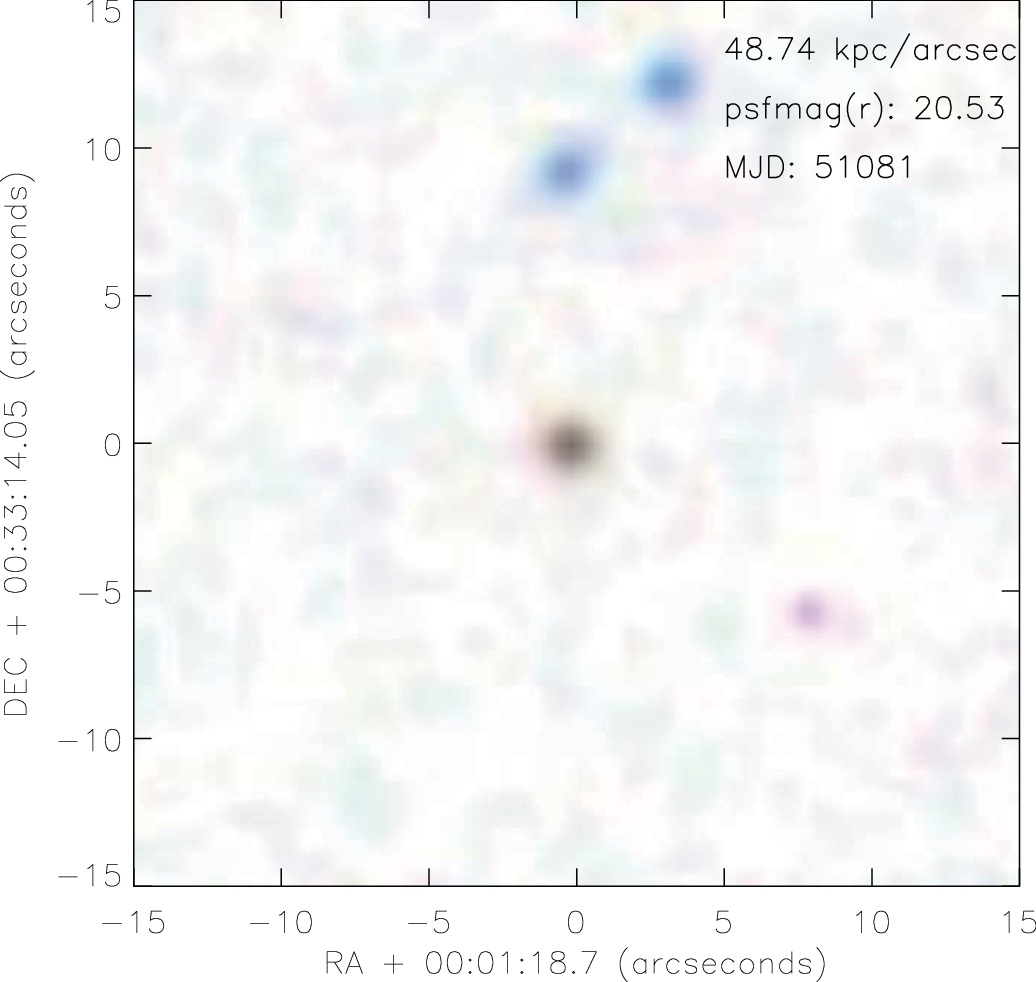}
\caption
{The inverted color image for the SDSS J0001 cut from the SDSS fits image (Flexible Image Transport System), which is constructed through the images of $gri$ bands. The kpc/arcsec scale and r band psfmag along with the MJD are shown in the top right region.}
\label{image}
\end{figure}

  TDEs in AGNs are valuable probes for the central SMBH and the origin of AGN variability. In general, a 
clear broad Mg~{\sc ii} emission line in the spectrum of a TDE candidate indicates that the host 
galaxy is an AGN. To date, only two events meeting the above criteria have been found. The 
first one is SDSS J120414.37+351800.5 in \citet{2018ApJ...859....8L}, though the quality of the light curve is quite limited, 
making it difficult to conduct further research. The second one was reported by \citet{2022MNRAS.516L..66Z}, who 
found a TDE candidate in SDSS J014124+010306 with broad Mg~{\sc ii} emission line. Motivated by this, we  
extensively searched for such candidates within the dataset of SDSS Stripe82 for quasars. We found one candidate 
in the quasar SDSS J000118.70+003314.0 (=SDSS J0001), characterized by broad Mg~{\sc ii} emission lines. This is the highest-redshift optical TDE candidate($\sim 1.404$) known in AGN. Figure \ref{image} shows a cutout of the image of SDSS J0001.

  In this manuscript, we report the long-term photometric SDSS $ugriz$-band variability of SDSS J0001 in 
Section 2. The theoretical TDE model and fitting procedure are given in Section 3 and Section 4, respectively. 
We provide our necessary discussion in Section 5. The summary and conclusions are shown in Section 6. Throughout 
the paper, $H_{0}$=70 km s$^{-1}$ Mpc$^{-1}$, $\Omega_{m}$=0.3, and $\Omega_{\Lambda}$=0.7 are adopted.

\section{Long-term light curves of SDSS J0001}
Detecting higher redshift TDEs as one of our recent objectives can provide further information on our understanding of the characteristics and evolution of TDEs around central SMBHs in high redshift galaxies and/or AGNs. 
Visually Inspecting the light curves of 7253 high redshift quasars with $z~>~1$ in SDSS Stripe8 reported in \citet{2010ApJ...721.1014M}, the light curves provided by the SDSS Stripe82 are checked one by one by eyes, leading to 
a smaple of $20$ candidates are obtained. Their light curves have a steep rise phase followed by a smooth decline. Among them, SDSS J0001 has the highest redshift. Therefore, the SDSS J0001 is of interest for our analysis. Detailed discussions on the small sample of TDE candidates in SDSS Stripe82 will be provided in our manuscript currently being prepared.

To construct the long-term variability curves in SDSS $ugriz$-bands of SDSS J0001, 
we collected optical light curves from the Stripe82 and the PHOTOOBJALL databases. SDSS 
Stripe82 is a region of the sky near the equator, covering about 300deg$^{2}$ area. It was imaged multiple times by the 
SDSS in u, g, r, i, and z bands from 2000 to 2008. They are 1.9-2.2 mag deeper than the best SDSS single-epoch data. The 
more detailed descriptions on the light curves by Stripe82 database \footnote{\url{https://das.sdss.org/va/stripe_82_variability/SDSS_82_public}}\citep{2008MNRAS.386..887B}. The PHOTOOBJALL database\footnote{\url{https://skyserver.sdss.org/dr16/en/help/browser/browser.aspx\#&&history=description+PhotoObjAll+U}}, the full photometric 
catalog quantities for SDSS imaging, contains one entry per detection, with the associated photometric parameters measured 
by PHOTO, and astrometrically and photometrically calibrated.

\begin{figure*}

\centering
    \includegraphics[width =0.33\linewidth]{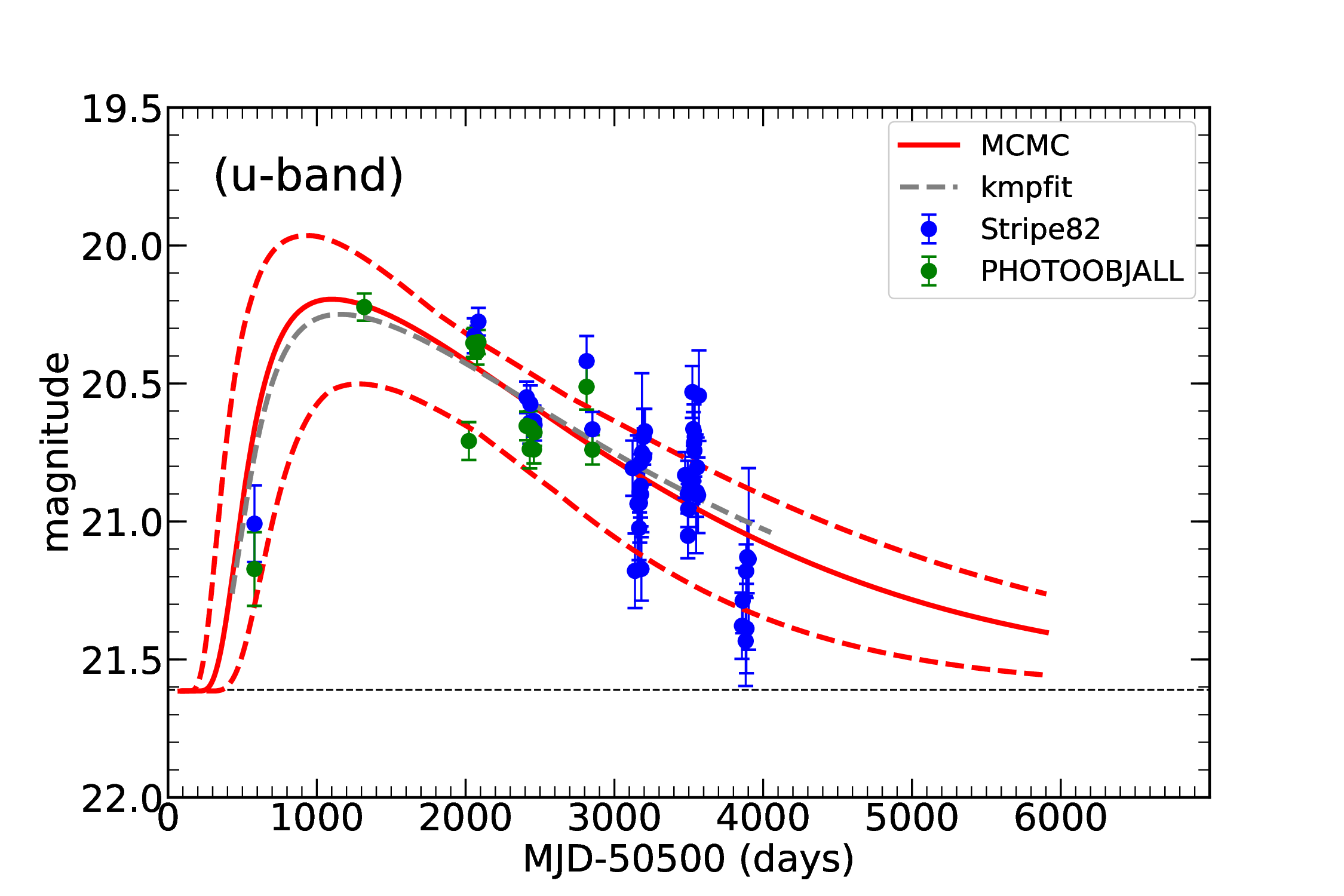}
    \includegraphics[width =0.33\linewidth]{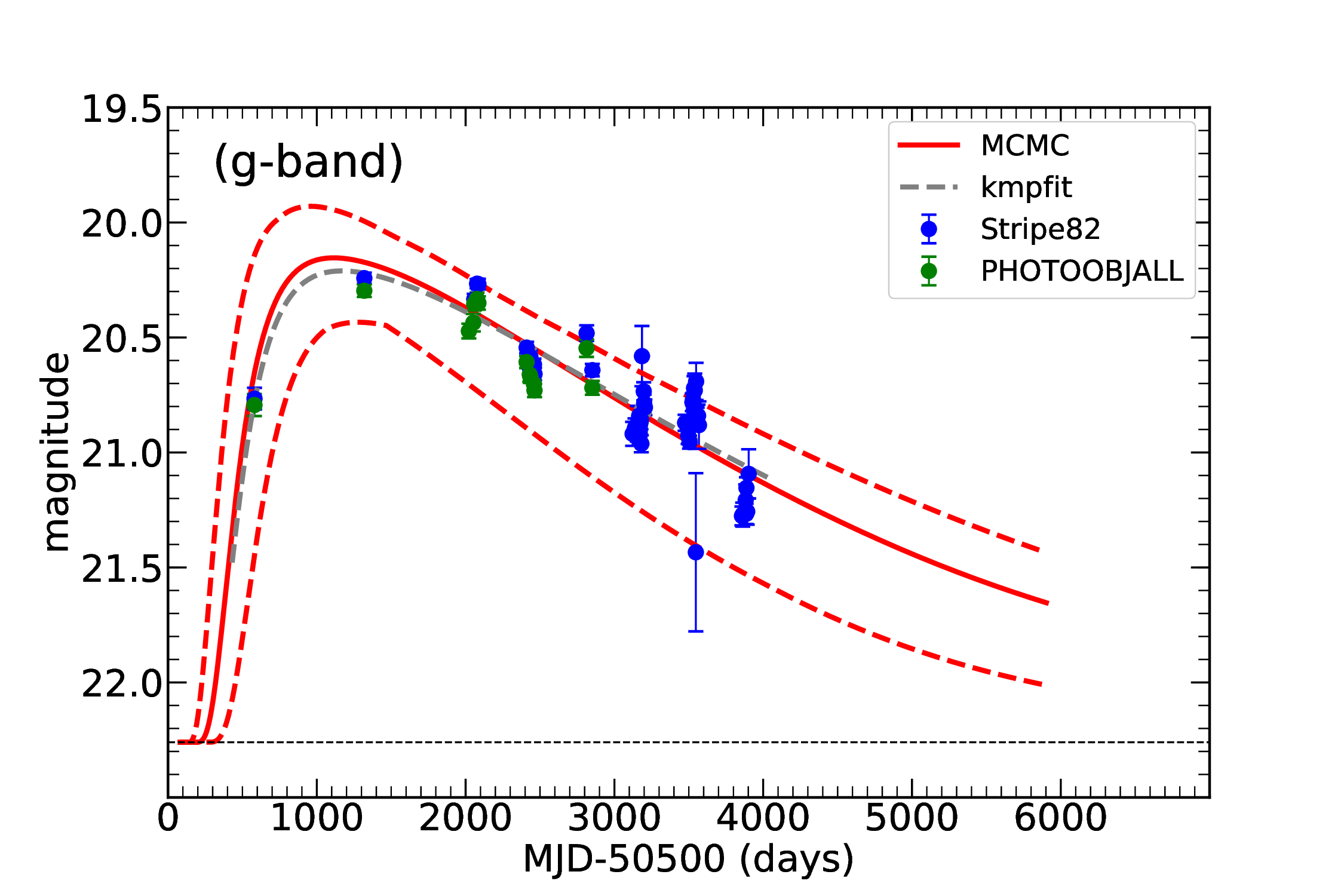}
    \includegraphics[width =0.33\linewidth]{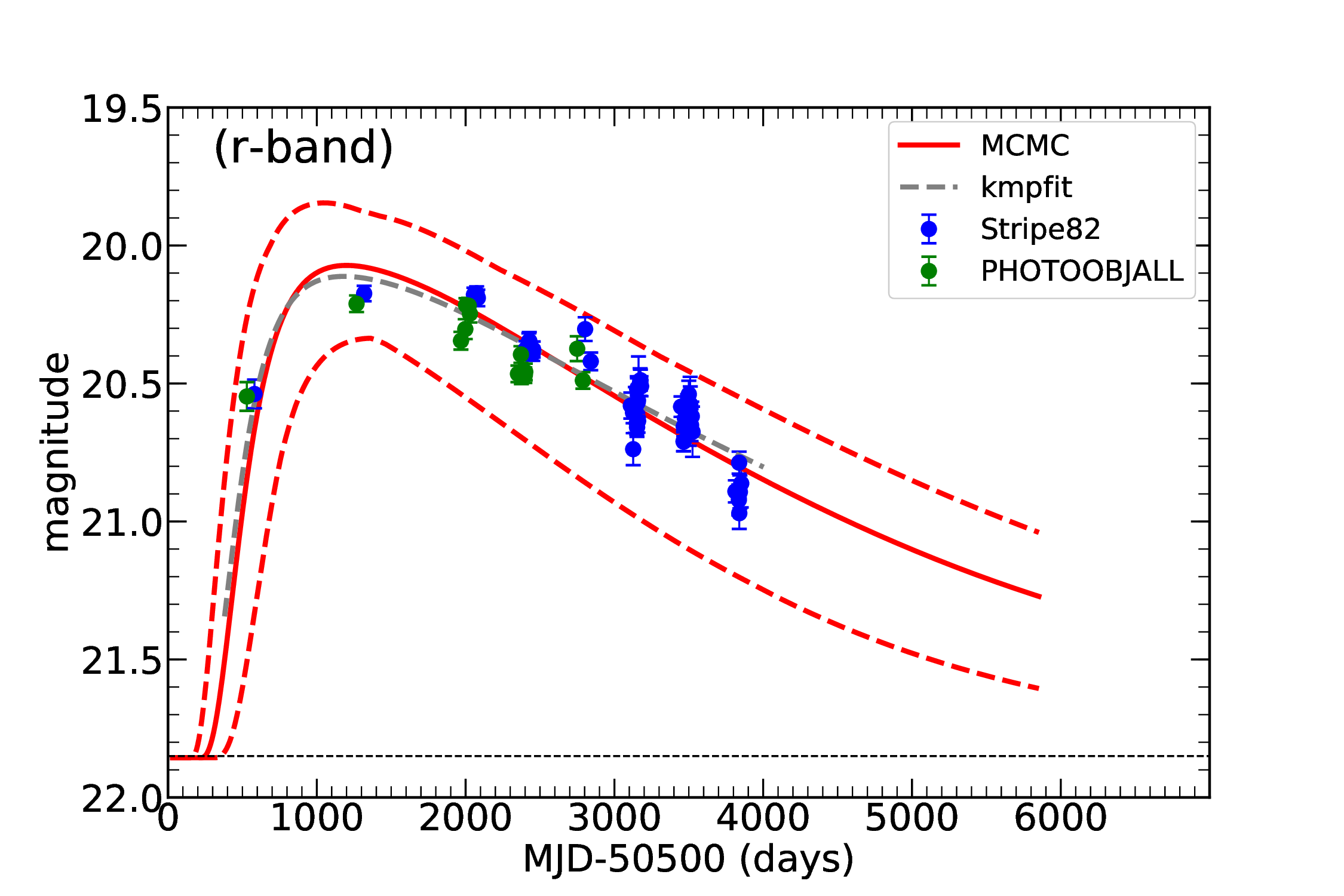}
    \includegraphics[width =0.33\linewidth]{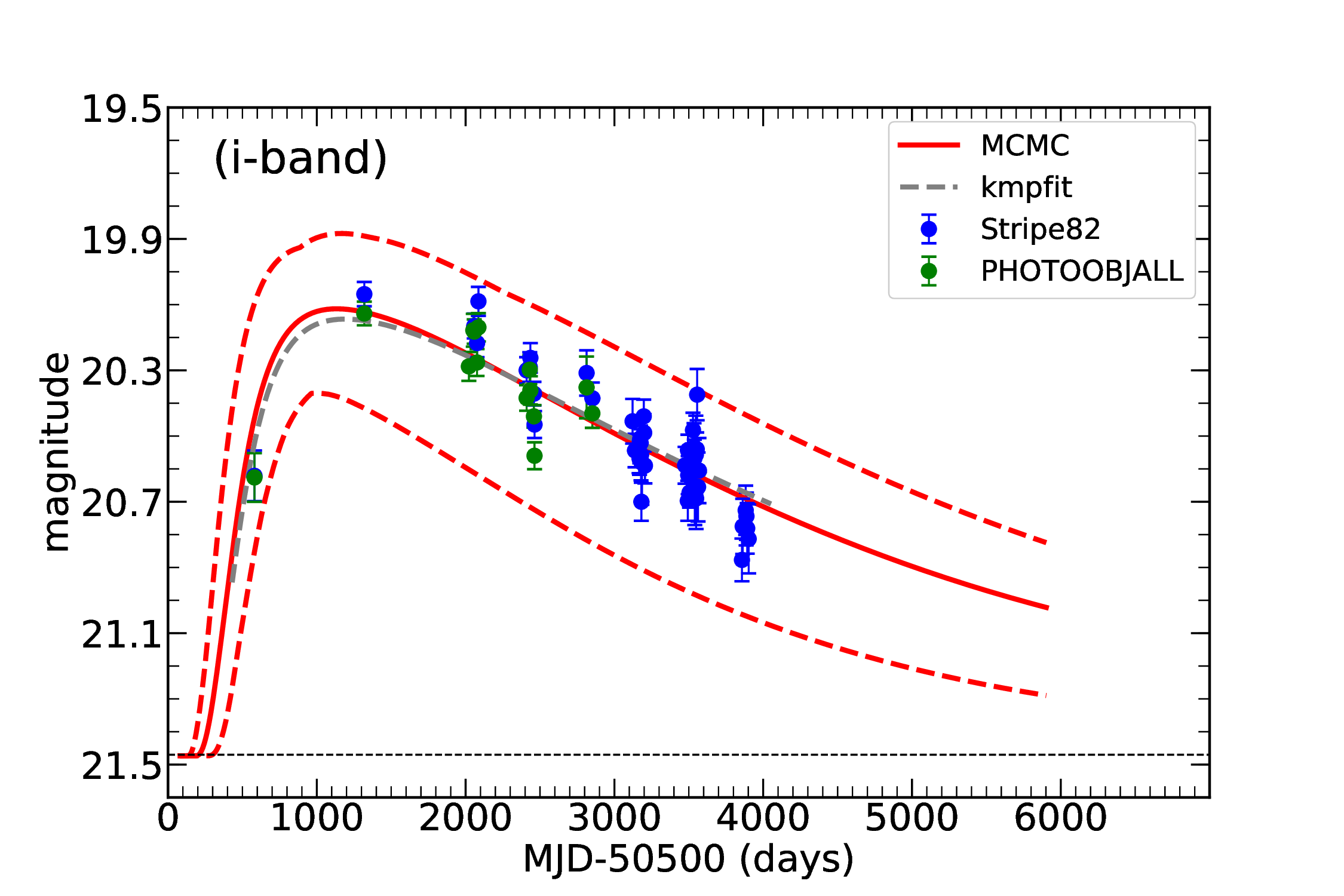}
    \includegraphics[width =0.33\linewidth]{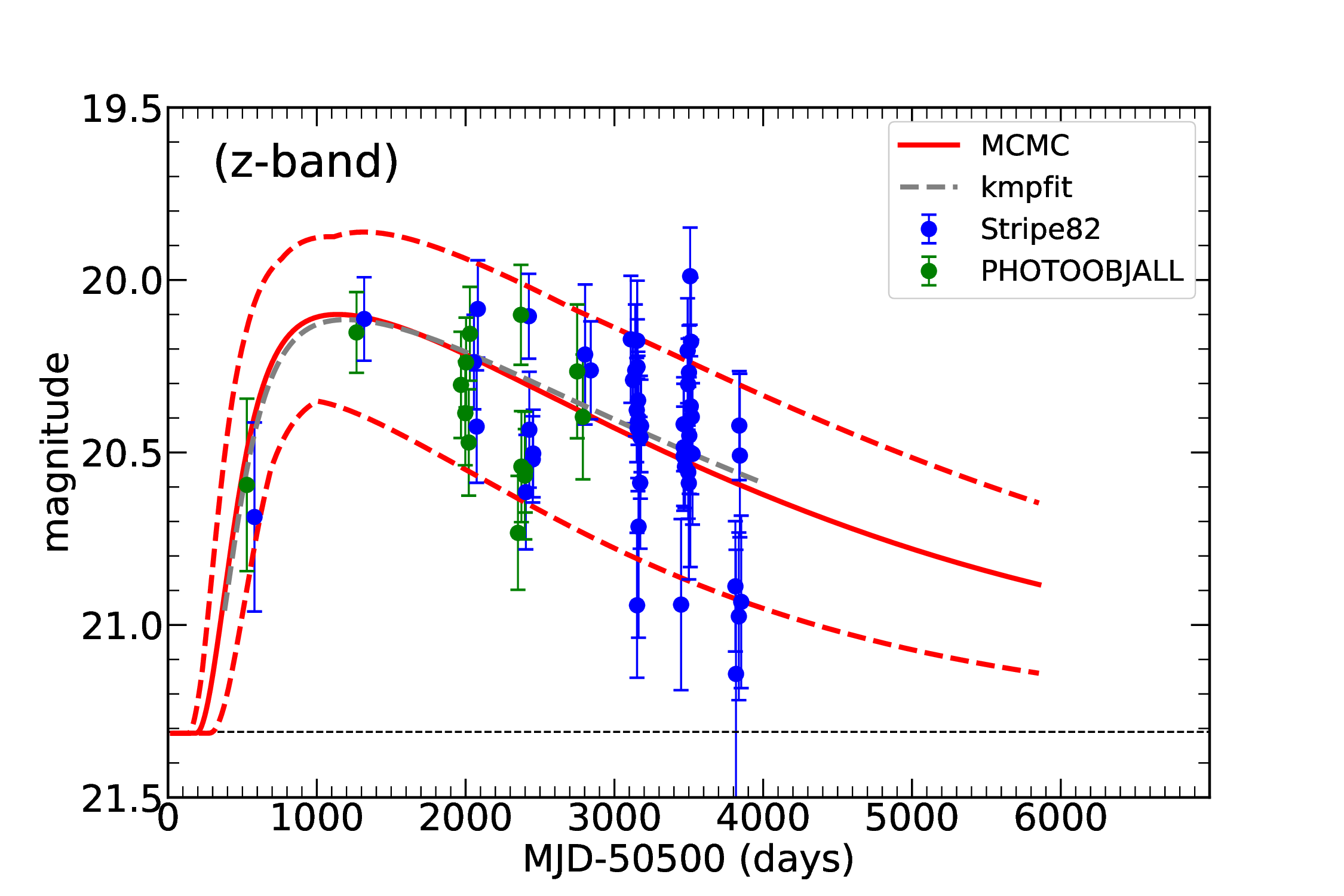}
    \includegraphics[width =0.33\linewidth]{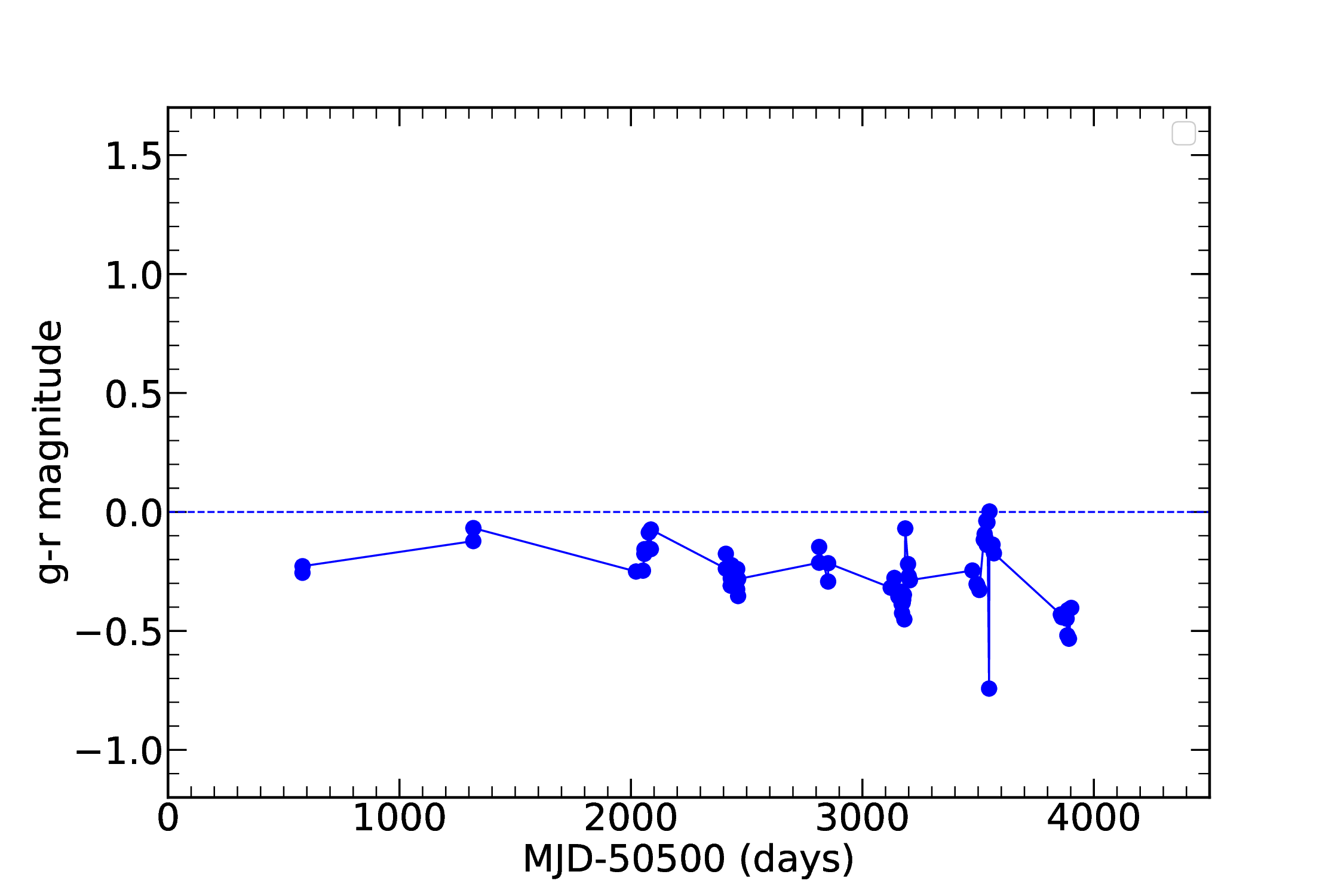}
 \caption
{Top panels and bottom left two panels show the observed $ugriz$ band light curves of SDSS J0001 and the best fit with the 
TDE model. Data points from both the Stripe82 and the PHOTOOBJALL databases have been used in our analysis. In each panel, solid blue and green circles with error bars represent the data points and the corresponding 
1$\sigma$ uncertainties from the Stripe82 and the PHOTOOBJALL databases, respectively. The grey dashed lines represent 
the fitting result by the kmpfit method. The red solid and dashed red lines show the best fit and the corresponding confidence 
bands determined by the 1$\sigma$ uncertainties of the model parameters with the MCMC technique, which is described and discussed in detail in Section \ref{sec 4}. The black dotted line represents 
the contributions from the host galaxy. Bottom right panel shows the color (g-r) evolution of SDSS J0001.}
 \label{1}
 
\end{figure*}

First, we collected the public light curves of SDSS J0001 from the Stripe82 database. They had a coverage of 9 years from 
1998 September 25 to 2007 October 29 (MJD 51082-54402). Second, the commonly accepted Structured Query Language (SQL) 
was applied to search for multi-epoch photometric light curves 
of SDSS J0001 in the PHOTOOBJALL database from data release 16 (DR16, \citet{2017AJ....154...28B, 2020ApJS..249....3A}), with 
the corresponding SQL query shown in the Appendix A. In this manuscript, only the data points are considered 
with magnitudes greater than 10 and less than 25 and positive uncertainties smaller than tenth of the corresponding magnitudes. 
Meanwhile, we compiled the \textit{ugriz}-band light curves from the PHOTOOBJALL, using the SDSS point-spread function (PSF) 
magnitudes because SDSS J0001 is a point-like source. The first five panels of Figure \ref{1} display the light curves in the 
\textit{ugriz}-band. In each panel, solid blue and green circles plus error bars represent the 
data points and the corresponding 1$\sigma$ uncertainties from the Stripe82 and the PHOTOOBJALL databases, respectively. 
The light curves with $\sim$ 60 data points in each band have the corresponding meantime step about 55 days.

The light curves of SDSS J0001 exhibit a clear rise-to-peak followed by a smooth decline trend, 
this unique variability is different from the continuous and long-term variability of EVQs (extreme variability
quasars) \citep{2018ApJ...854..160R,2024ApJ...963....7R}, resembling the variability behavior of common broad line AGN but with larger variability amplitudes. In other words, we don’t exclude the possibility that SDSS J0001 is an EVQ, but we expect a TDE to explain the unique variability in the light curve.

\begin{figure*}
\centering
\includegraphics[width = 6cm,height=5cm]{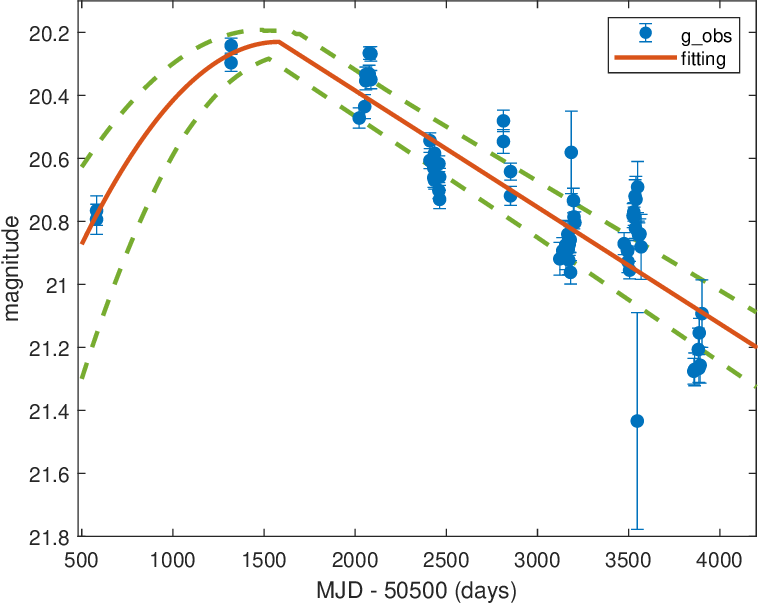}
\hspace {2.5cm}
\includegraphics[width = 6cm,height=5cm]{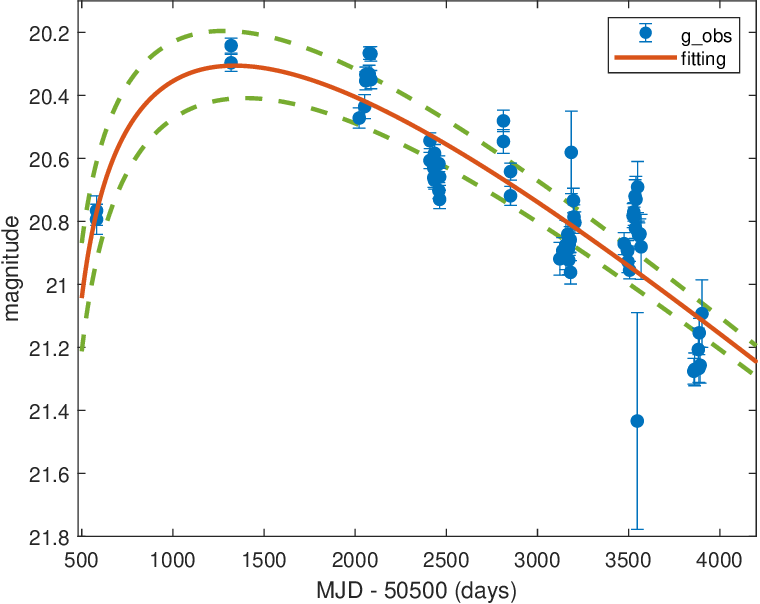}
\caption
{Best fit (black solid line) to the SDSS g-band light curve of SDSS J0001 (solid blue circles), with a 
Gaussian rise and an exponential decay (left panel) and Webull distribution (right panel). The red solid and dashed green lines show the best fit and the corresponding confidence bands determined by the 1$\sigma$ uncertainties of the model parameters with the Least Squares Method, respectively.}
\label{2}
\end{figure*}

To further analyze the characteristics of the light curves of SDSS J0001, we also 
characterized the light curves by applying other phenomenological models. Following \citet{2021ApJ...908....4V}, 
we modeled the light curves with a Gaussian rise and an exponential decay. The fitting results to the $g$-band light curve of 
SDSS J0001 are shown in the left panel of Figure \ref{2}, with fitting parameters $\log(\sigma)$=$2.99_{-0.14}^{+0.20}$ days and $\log(\tau)$=$3.46_{-1.19}^{+0.29}$ days, respectively. The corresponding errors are the 1$\sigma$ uncertainties determined by the Least Squares Method. The parameters $\log(\sigma)$ and $\log(\tau)$ of SDSS J0001, due to the longer 
duration of light curve, undoubtedly exceed those reported in \citet{2021ApJ...908....4V} for TDEs. However, the $\sigma/\tau$ $\sim$ 0.86
remains within their reported range (0.04-0.90). Following \citet{2021ApJ...908....4V}, the peak bolometric luminosity, photosphere temperature and radius are \( L = 2.28_{-0.51}^{+0.66} \times 10^{45} \, \text{erg/s} \), \( T = 2.64_{-0.05}^{+0.04} \times 10^4 \, \text{K} \), \( R = 2.56_{-0.39}^{+0.33} \times 10^{15} \, \text{cm} \), respectively. Compared to the TDE candidates reported in \citet{2019ApJ...872..151M,2021ApJ...908....4V,2023ApJ...942....9H}, the $L$, $R$, $T$ of SDSS J0001 are moderate among these TDE candidates.

Additionally, following 
\citet{2017MNRAS.470.4112G}, we further characterized the SDSS J0001 flare based on shape parameters utilizing the Weibull 
distribution. The Weibull shape parameter to the $g$-band light curve of SDSS J0001 is represented by $\log(a)$=$0.016_{-0.001}^{+0.002}$, and 
the scale parameter is denoted by $\log(s)$=$0.016_{-0.02}^{+0.02}$, respectively. The fitting results are shown in right panel of Figure 
\ref{2}. Notably, the parameter $\log(a)$ and $\log(s)$ of SDSS J0001 deviate completely from the parameter distribution of 
simulated flares according to single-point single-lens model(1S1L) and lensing candidates (see details in Fig 15 of \cite{2017MNRAS.470.4112G}). A small value of "a" indicates low symmetry, consistent with the flare characteristics of TDE. 
Additionally, the g-r color evolutions of 
SDSS J0001 (as shown in the bottom right panel of Figure \ref{1}) closely resemble the evolutionary characteristics
of TDEs described in \citet{2023ApJ...955L...6Y}.

\section{Theoretical TDE MODEL}

Based on the discussions above, the theoretical TDE model can be considered to describe the long-term 
variability of SDSS J0001 shown in the first five panels of Figure~\ref{1}. The details of the theoretical TDE model please refer to 
\cite{2013ApJ...767...25G}, \cite{2014ApJ...783...23G,2018ApJS..236....6G}, and \cite{2019ApJ...872..151M}. We briefly 
describe the model below.

\subsection{Dynamic Model}

	Assuming a TDE occurs when a star with mass $M_{*}$ is tidally disrupted by an SMBH with a mass of $M_{\rm BH}$, the 
template of fallback material rate $\dot{M}_{\rm fbt}$ evolves as
\begin{equation}
%\dot{M}_{fbt} = \frac{dM}{dE}\frac{dE}{dt} = \frac{2\pi}{3} \left(GM_{\rm BH,~6}\right)^{2/3} \frac{dM}{dE} t^{-5/3},
\dot{M}_{\rm {fbt}} = \frac{dM}{dE}\frac{dE}{dt}~ \sim ~\frac{(2~\pi~G~M_{\rm BH})^{2/3}}{3~t^{5/3}}\frac{dM}{dE},
\end{equation}
where $dM$ represents the debris mass, $dE$ is the specific binding energy after a star is torn apart, 
and $G$ is the gravitational constant, $dM/dE$ as distribution of debris mass on binding energy 
can be provided by TDEFIT/MOSFIT based on hydrodynamical simulations in \citet{2014ApJ...783...23G, 
2017ascl.soft10006G, 2018ApJS..236....6G}. Considering the viscous delay effects discussed in \citet{2013ApJ...767...25G, 
2019ApJ...872..151M}, the viscous-delayed accretion rates $\dot{M}_{a}$ from the fallback material rate $\dot{M}_ {\rm fbt}$ 
are given by
\begin{equation}
\dot{M}_{a}(T_{v}, \beta) ~=~\frac{exp(-t/T_{v})}{T_{v}}\int_{0}^{t}exp(t'/T_{v})\dot{M}_{\rm fbt}dt',
\end{equation}
where $T_{v}$ is the viscous time. \cite{2014ApJ...783...23G, 2018ApJS..236....6G} and \cite{2019ApJ...872..151M} 
developed the TDEFIT/MOSFIT code through hydrodynamical simulations for a standard TDE case with a star with mass 
of $M_{*}=1{\rm M_\odot}$ disrupted by a SMBH with mass of $M_{\rm BH,6}=1$, where $M_{\rm BH,6}$ represents central BH mass in units of $10^{6}{\rm M}_\odot$. The code presents the 
$dM/dE$ values by varying the impact parameter $\beta^{\rm temp}$ for different values, and considering a polytropic index of
$\gamma=4/3$ or $\gamma=5/3$.   

\begin{figure*}
\centering
\includegraphics[width = 7cm,height=5cm]{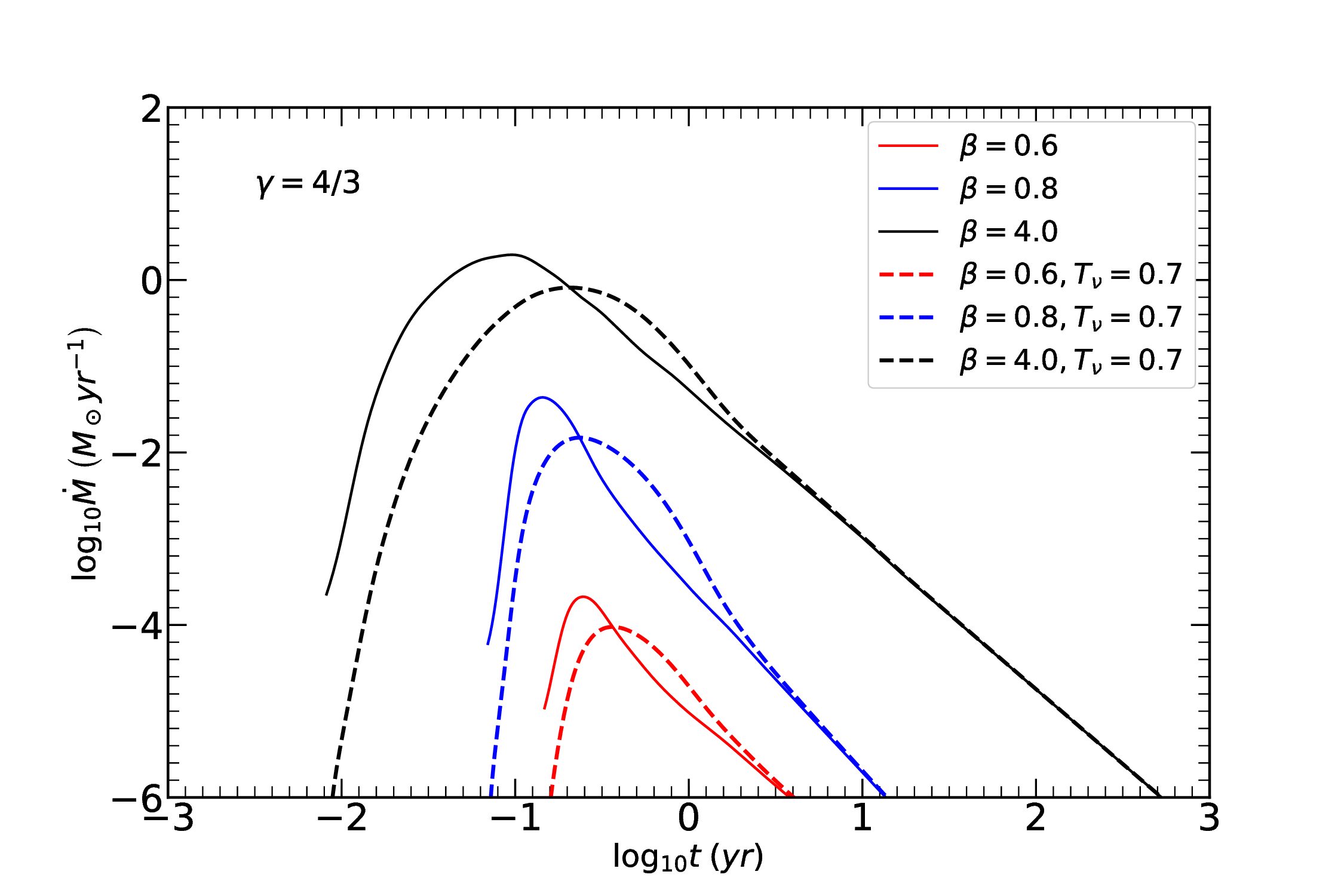}
\includegraphics[width = 7cm,height=5cm]{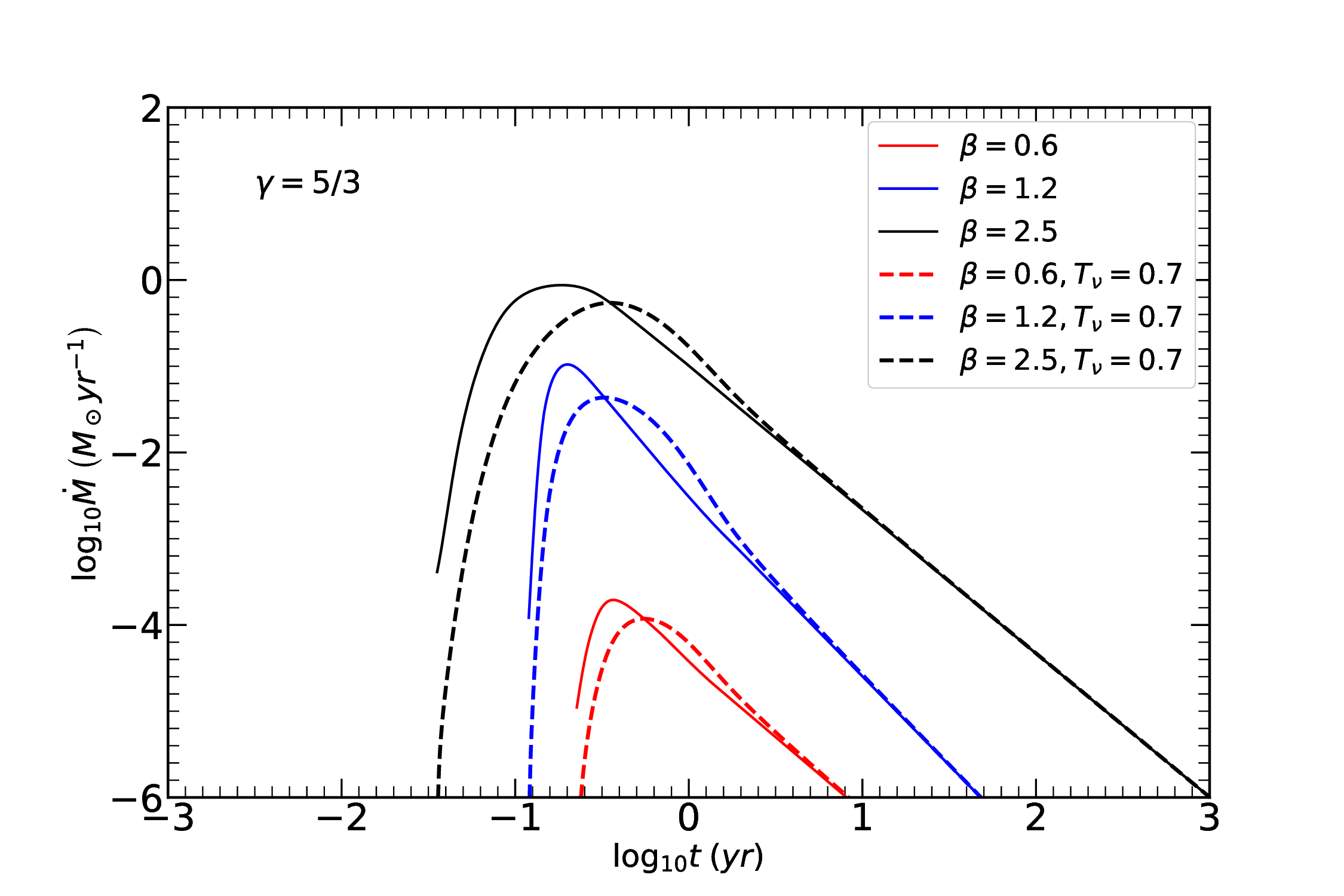}
\caption{The fallback material rate ($\dot{M}_{\rm fbt}$) (solid lines) and viscous-delayed accretion rate ($\dot{M}_{\rm at}$ with $T_v=0.7$ as an example) (dashed lines) of a $1{\rm M_\odot}$ star disrupted by $10^6{\rm M_\odot}$ black hole determined through the theoretical TDE model with polytropic index $\gamma=4/3$ (left panel) and $\gamma=5/3$ (right panel), respectively. As shown in the legend in each panel, line styles in different colors represent the results with different $\beta$.}
\label{10}
\end{figure*}

The numerical results then can be taken as a template for deriving the $\dot{M}_{a}(T_{v}, \beta)$ with $\beta$ 
different from $\beta^{\rm temp}$ through interpolations. We collect the $dM/dE$ values from the TDEFIT/MOSFIT 
code by varying the impact parameters value $\beta^{\rm temp}$  from 0.6 to 4.0 for the scenario of $\gamma=4/3$ (23 values) 
or from 0.5 to 2.5 for $\gamma=5/3$ (20 values) and linear varying $\log T_{v}^{temp}$ from -3 to 0 ($31$ values). Hence, we 
construct templates of the time-dependent viscous-delayed accretion rates $\dot{M}^{\rm temp}_{a}$ of 713 (620) viscous-delayed 
accretion rates for polytropic index $\gamma=4/3$ (or $\gamma=5/3$).  Figure \ref{10} illustrate temporal evolution of $\dot{M}_{\rm fbt}$ and $\dot{M}_{a}$ for some sets of $\{T_{v}, \beta\}$ as marked in the figure. One can observe that $\dot{M}_{a}$ as a function of time peaks at a later time and has a lower peak value than that of $\dot{M}_{\rm fbt}$. This is reasonable regarding the mass reservation of the central accretion disk.

We calculate the viscous-delayed accretion rate $\dot{M}_{a}(T_{v},\beta)$ for a given set of $\{T_v,\beta\}$  
through linear interpolations. The first linear interpolations are applied to obtain the viscous-delayed 
accretion rates for the input $T_{v}$ using the following two equations
\begin{equation}
\begin{split}
\dot{M}_{a}(T_{v}, \beta^{\rm temp}_1) &= \dot{M}^{\rm temp}_{a}(T^{\rm temp}_{v1}, \beta^{\rm temp}_1) + 
\frac{T_{v}-T^{\rm temp}_{v1}}{T^{\rm temp}_{v2}-T^{\rm temp}_{v1}}\\
&[\dot{M}^{\rm temp}_{a}(T^{\rm temp}_{v2}, \beta^{\rm temp}_1) 
	- \dot{M}^{\rm temp}_{a}(T^{\rm temp}_{v1}, \beta^{\rm temp}_1)],\\
\dot{M}_{a}(T_{v}, \beta^{\rm temp}_2) &= \dot{M}^{\rm temp}_{a}(T^{\rm temp}_{v1}, \beta^{\rm temp}_2) + 
\frac{T_{v}-T^{\rm temp}_{v1}}{T^{\rm temp}_{v2}-T^{\rm temp}_{v1}}\\&[\dot{M}^{\rm temp}_{a}(T^{\rm temp}_{v2}, \beta^{\rm temp}_2)
	- \dot{M}^{\rm temp}_{a}(T^{\rm temp}_{v1}, \beta^{\rm temp}_2)],
\end{split}
\end{equation}
where $\beta^{\rm temp}_1$ ($T^{\rm temp}_{v1}$) and $\beta^{\rm temp}_ 2$ ($T^{\rm temp}_{v2})$ are two 
adjacent template $\beta$ ($T^{\rm temp}_{v}$) values that satisfy $\beta^{\rm temp}_1\le\beta\le\beta^{\rm temp}_ 2$  
($T^{\rm temp}_{v1}\le T_{v}\le T^{\rm temp}_{v2}$). Then the viscous-delayed accretion rates for the input $T_{v}$ and 
$\beta$ can be obtained by 
\begin{equation}
\begin{split}
\dot{M}_{a}(T_{v}, \beta) = \dot{M}_{a}(T_{v}, \beta^{\rm temp}_1) + 
 	\frac{\beta-\beta^{\rm temp}_1}{\beta^{\rm temp}_2-\beta^{\rm temp}_1}\\
[\dot{M}_{a}(T_{v}, \beta^{\rm temp}_2)- \dot{M}_{a}(T_{v}, \beta^{\rm temp}_1)].
\end{split}
\end{equation}
We show an example in Figure \ref{int} of results through these three linear interpolation processes in the Appendix B.

For a general TDE with BH mass $M_{\rm BH}$ and star mass $M_{*}$ in units of ${\rm M_\odot}$, we calculate the values of $\dot{M}$ as a function of time by adopting the following scaling relations \citep{2019ApJ...872..151M},
\begin{equation}
\begin{split}
&\dot{M} = M_{\rm BH,6}^{-0.5}\times M_{*}^2\times
	R_{*}^{-1.5}\times\dot{M}_{a}(T_{v}, \beta) \\
&t = (1+z)\times M_{\rm BH,6}^{0.5}\times M_{*}^{-1}\times
	R_{*}^{1.5} \times t_{a}(T_{v}, \beta),
\end{split}
\end{equation}
where $R_{*}$ represents the stellar radius in units of ${\rm R_{\odot}}$ and $z$ is the redshift of the host galaxy 
of the TDE. Additionally, we adopt the known mass-radius relation for main sequence stars in \cite{1996MNRAS.281..257T}.

\subsection{Radiation Model}

The radiating region is assumed to be a standard blackbody photosphere, as discussed in \cite{2019ApJ...872..151M}. 
The time-dependent effective blackbody temperature $T_p(t)$ is estimated with  
\begin{equation}
\begin{split}
\begin{aligned}
&T_p(t) = \left (\frac{L}{4\pi\sigma_{\rm SB}R_p^2}\right)^{1/4} = 
\left[ \frac{\eta\dot{M(t)}c^2}{4\pi\sigma_{\rm SB}R_p^2}\right ]^{1/4},
\end{aligned}
\end{split}
\end{equation}
where $L$ is the time-dependent bolometric luminosity given by  $L = {\eta\dot{M(t)}c^2}$, $\eta$ represents 
the energy transfer efficiency that is lower than 0.4 \citep{2014ApJ...783...23G,2019ApJ...872..151M}, 
$\sigma_{\rm SB}$ is the Stefan-Boltzmann constant, $c$ is the speed of light, and $R_p(t)$ is the radius 
of the photosphere. $R_p(t)$  ranges from the minimum $R_{\rm isco}$ (event horizon radius) to the maximum 
semimajor axis ($a_p$) of the accreting mass, it is assumed to be a power-law dependence on $L$ (e.g. 
\citealt{2019ApJ...872..151M}), i.e. 
\begin{equation}
\begin{aligned}
&R_p(t) = R_0\times a_p(L/L_{\rm Edd})^{l_p} = R_0\times a_p\left [\frac{\eta\dot{M(t)}c^2}{1.3\times10^{38}M_{\rm BH}}\right ]^{l_p},
\end{aligned}
\end{equation}
where $L_{\rm Edd}$ represents the Eddington luminosity (${L_{{\text{Edd}}}} \equiv 4G{M_{\text{BH}}}c/\kappa$  
and $\kappa$ is the mean opacity to Thomson scattering assuming solar metallicity), $R_{0}$ is a dimensionless 
free parameter, $l_p$ represents the power-law exponent and $t_p$ is time of the peak accretion rate. 
The value of $a_p$ is given by
\begin{equation}
\begin{aligned}
&a_p = \left [G M_{\rm BH}\times (\frac{t_p}{2\pi})^2\right]^{1/3}.	
\end{aligned}
\end{equation}
The time-dependent emission spectrum in the rest frame can be calculated as
\begin{equation}
\begin{split}
&F_\lambda(t)=\frac{2\pi hc^2}{\lambda^5}\frac{1}{e^{hc/(k\lambda T_p(t))}-1}\left[\frac{R_p(t)}{D(z)}\right]^2,\\
\end{split}
\end{equation}
where $D(z)$ is the luminosity distance at redshift $z$. 

After calculating the time-dependent $F_\lambda(t)$ in the observer frame and then convolved with the 
transmission curves of the SDSS $ugriz$ filters, the time-dependent apparent magnitudes $mag_{u,~g,~r,~i,~z}(t)$ 
can be determined in the corresponding SDSS bands. Then, we can check whether the MOSFIT TDE model can 
be applied to describe the long-term variability of SDSS J0001 shown in the first five panels of Figure~\ref{1}.

\section{Fitting procedure and results }\label{sec 4}
We fit the light curves in the $ugriz$ bands of SDSS J0001 by the MOSFIT TDE model. The free parameters of the model include $M_{\rm BH}$, $M_{*}$, $R_{*}$, $\beta$, $T_{v}$, $\eta$, $R_0$, and $l_p$. The brightness [$mag_{0}(u,~g,~r,~i,~z)$] of the host galaxy is taken as free parameter. The model requires that the tidal disruption radius $R_{\rm TDE}$ derived from a set of model parameters is larger than the event horizon of the central BH. 
In order to fit the observed light curves, the following two steps are applied. 
First, we fit the light curves with the Levenberg-Marquardt least-squares optimization technique (the kmpfit module in PYTHON; \citealp{2009ASPC..411..251M}). Being due to the poorly light curve sampling, the uncertainties of the model parameter cannot be constrained with the kmpfit module. Thus, we secondly adopt the Markov Chain Monte Carlo (MCMC) technique (the emcee package in PYTHON) \citep{2013PASP..125..306F} to improve our fit. The prior model distributions, listed in Table 1, are set based on the results of the kmpfit module. 400 MCMC iterations with 500 walkers have been applied in our MCMC fit. 

\begin{figure*}
\centering
\includegraphics[width=1\linewidth]{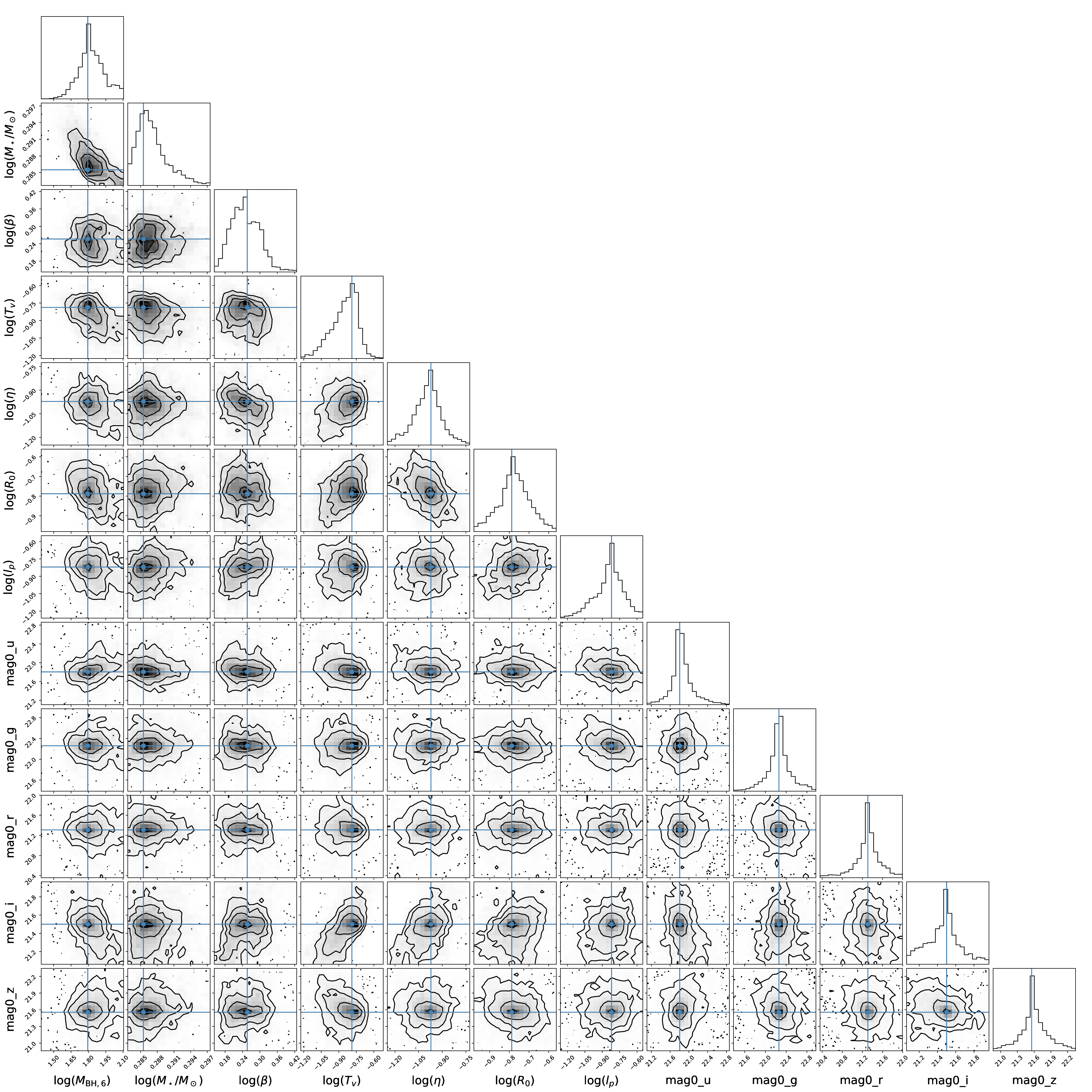}
\caption
{Corner plot shows the posterior probability distributions of the TDE model parameters obtained from the 
MCMC technique. In each panel, the three circles from outer to inner represent $3\sigma$, $2\sigma$, and 
$1\sigma$ confidence levels, and the blue dot in the center of each contour marks the position of the 
best-fit parameter.}
\label{4}
\end{figure*}

Our fitting results are shown in the first five panels of Figure~\ref{1}. The derived posterior distributions of the model parameters are shown in Figure~\ref{4}, and best fit parameters are reported in Table 1. The reduced $\chi^2/dof$ of our fit is $\sim4.5$. One can observe that the light curves of SDSS J0001 can be explained as a tidal disruption of a main sequence star with mass of $1.905_{-0.009}^{+0.023}{\rm M_\odot}$ by a SMBH with mass of $M_{\rm BH}\sim6.5_{-2.6}^{+3.5}\times10^7{\rm M_\odot}$. The predicted peak brightness of the TDE event is at $1.83_{-0.33}^{+0.46}$ years. The total energy of the event derived from our fit is $2.38\times10^{53}$ ergs and about 0.78 ${\rm M_\odot}$ of debris mass is accreted by the central SMBH (corresponding 1.12 ${\rm M_\odot}$ of the disrupted star are ejected). 

\begin{figure*}
\centering
\includegraphics[width = 6cm,height=5cm]{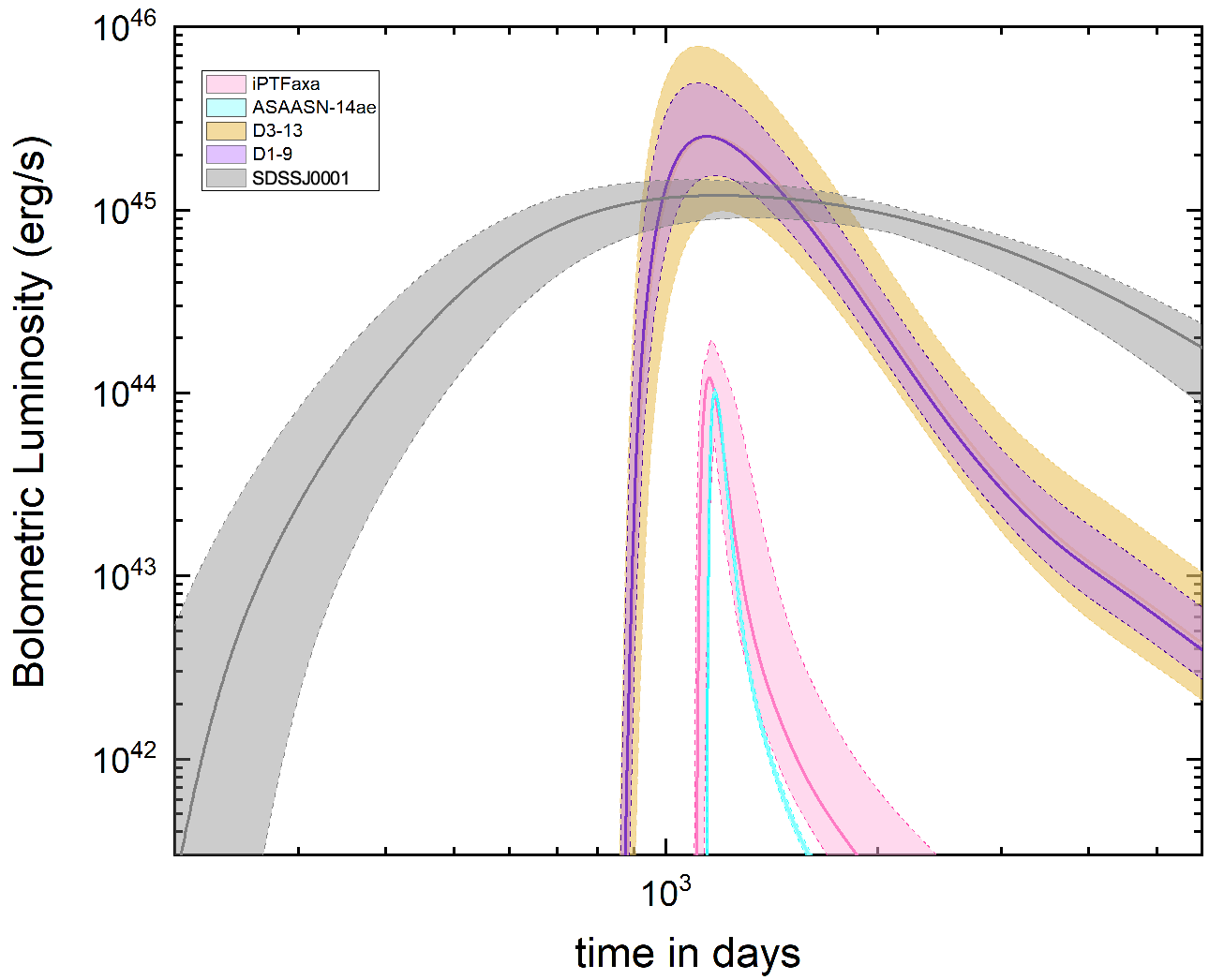}
\hspace{2.5cm}
\includegraphics[width = 6cm,height=5cm]{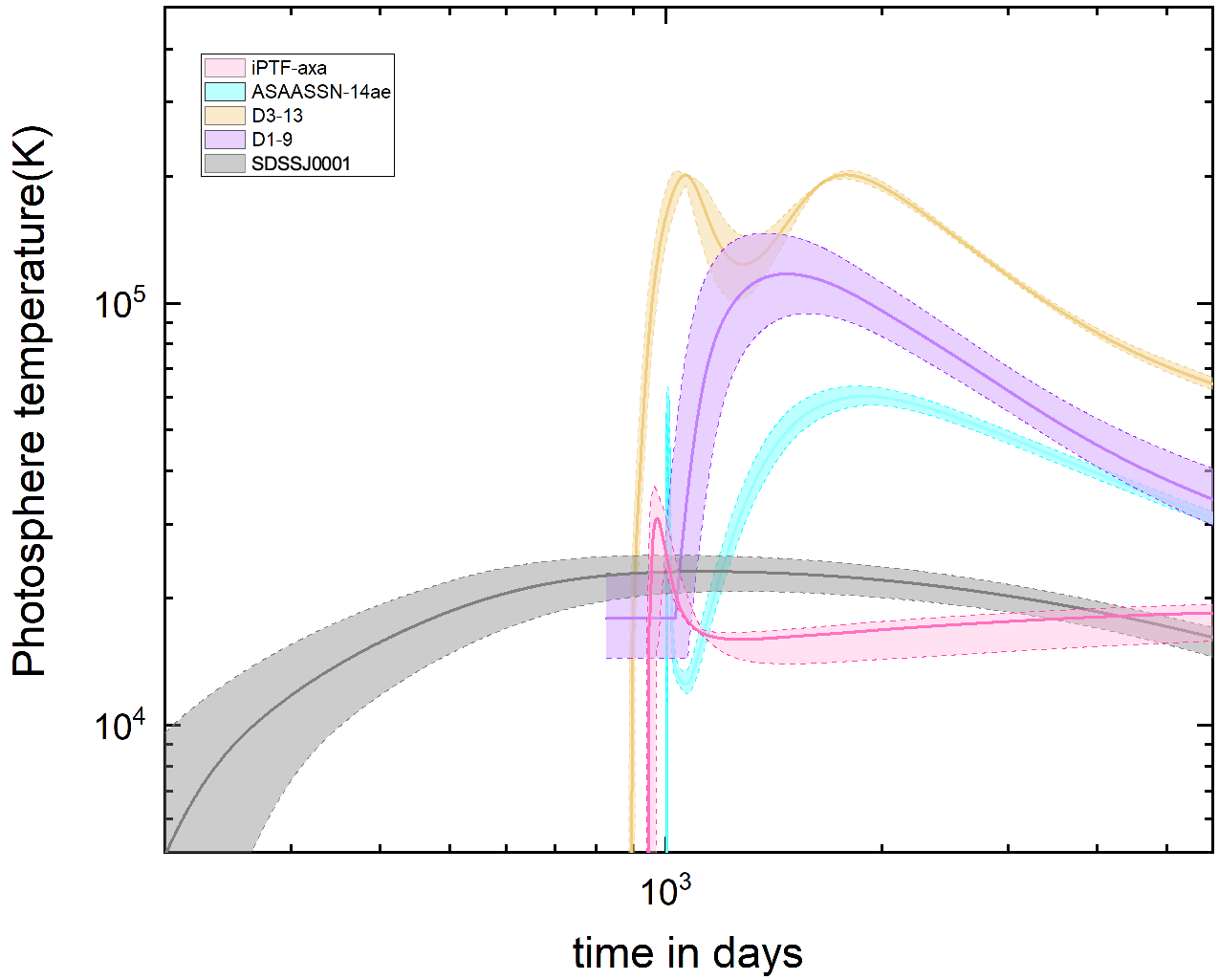}
\caption
{Temporal Evolution of the bolometric luminosity ({\em left panel}) and photosphere temperature ({\em right panel}) of SDSS J0001 in comparison with that of four events from \citet{2019ApJ...872..151M} as marked in each panel with different colors. All events 
align with their peak bolometric luminosity or photosphere temperature, and the time is in the observer's frame. The shaded regions indicate 
the confidence bands determined by uncertainties of model parameters.}
\label{6}
\end{figure*}

The model predicted time-dependent bolometric luminosity and photosphere temperature of SDSS J0001 are shown in Figure~\ref{6}. The peak bolometric luminosity and photosphere temperature are $1.18_{-0.31}^{+0.26} \times 10^{45}$ erg/s and $2.30_{-0.28}^{+0.30}\times 10^{4}$ K, respectively. Adopting the parameters of four TDEs reported by  \citet{2019ApJ...872..151M}, we also compare the time-dependent bolometric luminosity and photosphere temperature of the four TDEs with SDSS J0001 in Figure~\ref{6}. 
%also shows the results of other four TDEs that are reported by \citet{2019ApJ...872..151M}. 
Among them    
the TDEs of D1-9 \citep{2008ApJ...676..944G} and D3-13 \citep{2008ApJ...676..944G} have a relatively high bolometric luminosity, and TDEs ASASSN-14ae
\citep{2014MNRAS.445.3263H,2016MNRAS.462.3993B} and iPTF-axa \citep{2017ApJ...842...29H} have relatively a low bolometric
luminosity. It is found that the intrinsic peak bolometric luminosity and photosphere temperature of  SDSS J0001 are moderate among these TDE candidates.

\begin{table}
	\centering
	\label{par_MC}
\caption{Parameters of the TDE model derived from our Kmpfit and MCMC fit to the light curves of SDSS J0001.}

 \begin{tabular}{lccr} % four columns, alignment for each
            \hline\hline
    Parameters & Prior distribution & Kmpfit & MCMC fit \\ 
        \hline
$\log(M_{\rm BH,~6})$ & [-1, 3] &1.79& $ 1.81_{~-0.22}^{~+0.19}$  \\
$\log(M_\star/M_\odot)$ & [-2, 1.7] & 0.28 & $ 0.28_{~-0.002}^{~+0.005}$ \\
$\log(\beta)(4/3)$ & [-0.22, 0.6] & 0.25 & $0.25_{~-0.063}^{~+0.084}$ \\
$\log(T_v)$ & [-3, 0] & -0.76 & $-0.84_{~-0.24}^{~+0.21}$ \\ 
$\log(\eta)$  & [-3, -0.4] &-0.97 & $-0.98_{~-0.13}^{~+0.15}$  \\
$\log(R_{0})$ &  [-3, 3] & -0.82 & $-0.77_{~-0.23}^{~+0.22}$  \\
$\log(l_{p})$  & [-3, 0.6] & -0.82 & $-0.82_{~-0.13}^{~+0.34}$  \\
$mag_{0}(u)$ & [20, 30] & 21.56 &$21.61_{~-0.56}^{~+0.72}$\\
$mag_{0}(g)$ & [20, 30] & 22.26 &$22.25_{~-0.53}^{~+0.49}$\\
$mag_{0}(r)$ & [20, 30] & 21.80 &$21.85_{~-0.43}^{~+0.60}$\\
$mag_{0}(i)$ & [20, 30] & 21.50 & $21.47_{~-0.35}^{~+0.62}$\\
$mag_{0}(z)$ & [20, 30] & 21.31 & $21.31_{~-0.59}^{~+0.57}$\\
        \hline\hline
	\end{tabular}

\end{table}

\section{DISCUSSIONS}

\subsection{Timescale of the TDE Candidate SDSS J0001}

\begin{figure*}
\centering
\includegraphics[width = 8cm,height=12cm]{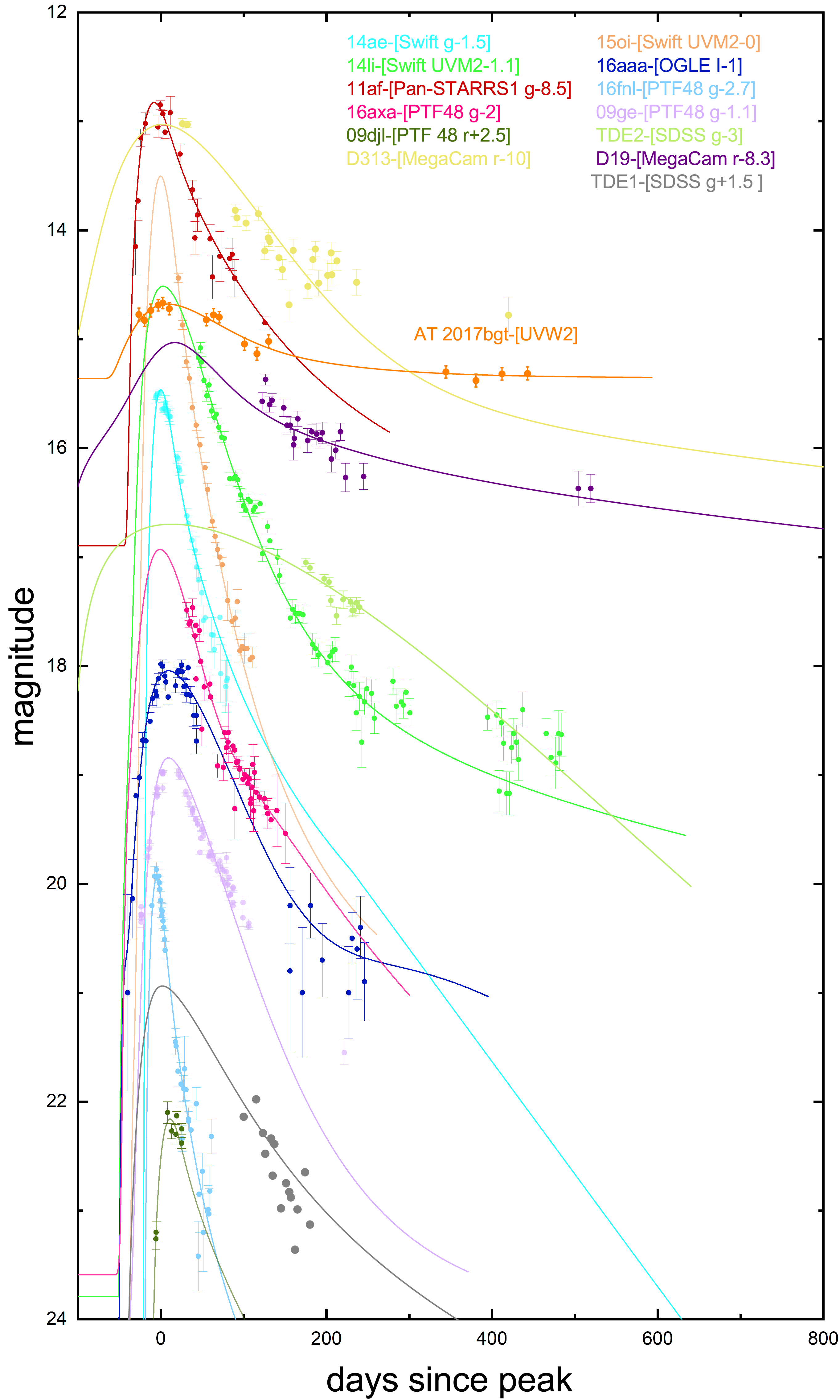}
\caption
{Comparison of the slowly-decayed UV-optical transient AT 2017bgt reported by \citet{2019NatAs...3..242T} with the 13 TDE candidates reported by \citet{2019ApJ...872..151M}. The best fit results by applying the fitting parameters from \citet{2019ApJ...872..151M} in our fitting procedure. Note that no theoretical light curve is available for the TDE candidate PS1-10jh because the estimated tidal disruption radius is smaller than the event horizon of the central BH.}
\label{12}
\end{figure*}

We collected light curves of the typical TDE candidates in \citet{2019ApJ...872..151M} and re-described them using their model parameters, as applied in our method, as shown in Figure \ref{12}\footnote{We collected the light curves of the 13 TDE candidates from \url{https://tde.space/.}}. Compared with these TDEs, SDSS J0001 has the longest timescale. Noting that $t\propto {M_{\text{BH}}}^{1/2} M_{*}^{-1}R_{*}^{1.5}$ and considering the mass-radius relation for main sequence stars, we have $t\propto {M_{\text{BH}}}^{1/2} M_{*}^{1/2}$. The long timescale is partially due to the large black hole mass and large stellar mass. More important, as shown in Figure \ref{10} (see also Fig 5 of \citet{2013ApJ...767...25G}), the timescale is sensitive to the impact parameter $\beta$. A larger $\beta$ leads to a longer timescale. The derived $\beta$ value for SDSS J0001 is 1.77, which is indeed much larger than other TDE candidates reported by \citet{2019ApJ...872..151M} except for D3-13. The $\beta$ value of TDE D3-13 is 1.8. Its timescale exceeds one thousand days. Considering the $M_{\rm BH}$ of SDSS J0001 and the time dilation effect of SDSS J0001 at high redshift ($z=1.4$), its intrinsic timescale is comparable to TDE D3-13.

  \cite{2019NatAs...3..242T} proposed that the slowly-decayed UV-optical transient AT 2017bgt is likely a new type of flare associated with accreting SMBHs, and suggested that it is unlikely to be a TDE driven flare. Inspecting the light curves of AT 2017bgt in the optical, UV, and X-ray bands reported by \cite{2019NatAs...3..242T}, the rise-to-peak and followed by smooth declining trend in UV band light curve is apparent and can be well expected by a central TDE. We compare it with other TDE candidates reported by \citet{2019ApJ...872..151M} in Figure \ref{12}. We also present the theoretical light curves with our TDE model code by adopting the parameters the same as that reported by \citet{2019ApJ...872..151M}. Using the kmpfit module, we fit its UVW1 and UVW2 light curves, with the UVW2 fitting curve is also shown in Figure \ref{12}. Our findings suggest that these curves are well represented by the TDE model
  with parameters of $M_*=7.76{\rm M_\odot}$, $M_{\rm BH}=2.0\times10^6{\rm M_\odot}$, $\beta=1.14$, and $\gamma=4/3$. We cannot entirely rule out the possibility that AT 2017bgt is a TDE candidate.\footnote{We should carefully note that the loss of data points around 200-350 days of AT2017bgt light curve makes uncertainty of the late decaying behavior of AT 2017bgt.}

\subsection{Host galaxy and SMBH mass}
\begin{figure*}
\centering
\includegraphics[width = 8cm,height=6cm]{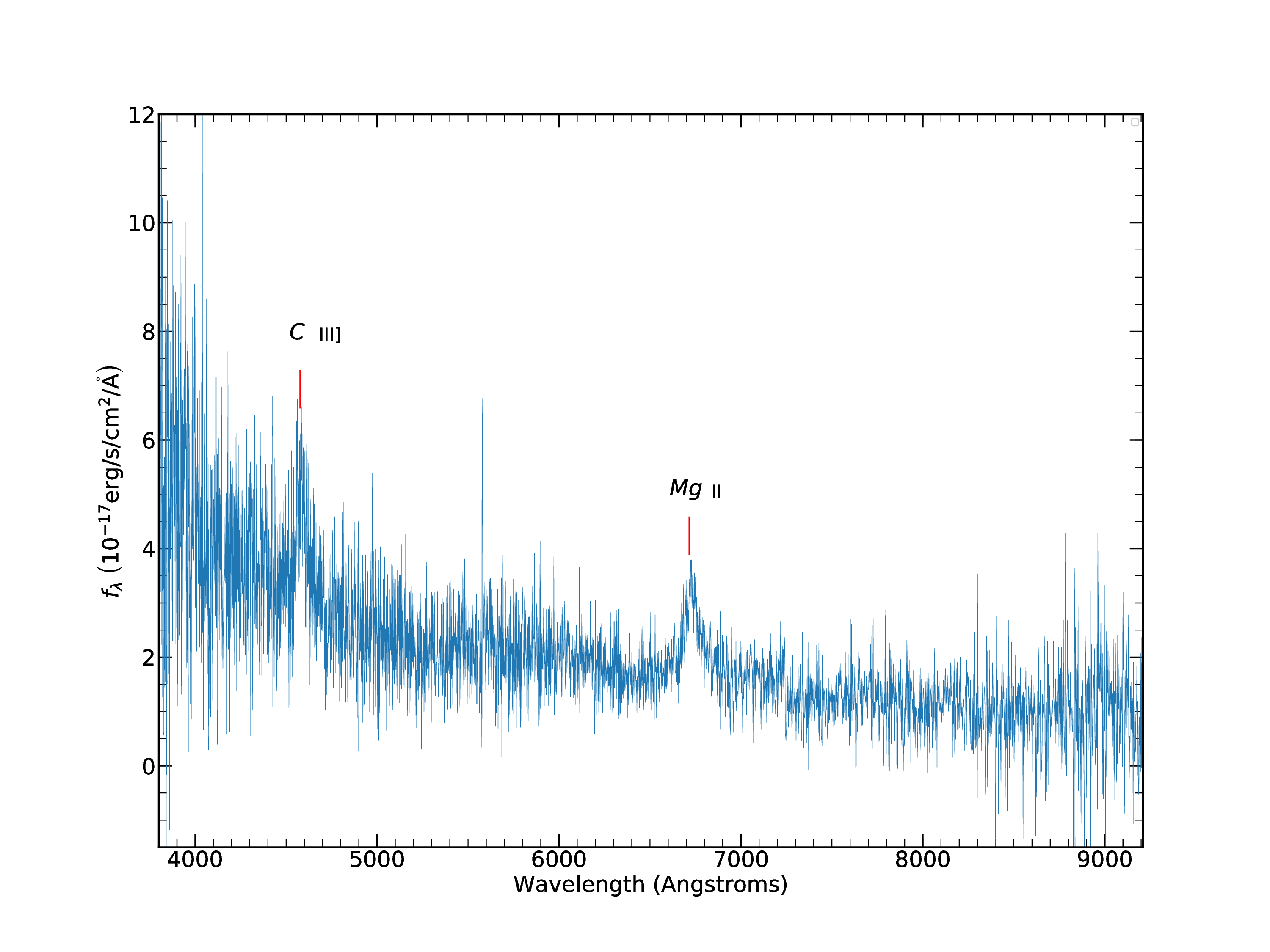}
\caption 
{Optical spectrum of SDSS J0001 in observer frame. The vertical solid red lines mark positions of the broad lines.}
\label{5}
\end{figure*}

The SDSS spectrum of SDSS J0001 (plate-mjd-fiberid = 1091-52902-525) observed around the epoch of peak brightness of its light curve, collected from SDSS DR16\citep{2020ApJS..249....3A}, is shown in Figure \ref{5}. It has a quasar-like spectrum with apparent broad Mg~{\sc ii} emission line. \citet{2011ApJS..194...45S} estimated its virial BH mass as $4.9\times10^8{\rm M_\odot}$ 
through broad Mg~{\sc ii} emission lines, which is approximately 7.5 times larger than that derived from our analysis. 
It is possible that the broad Mg~{\sc ii} emission materials in SDSS J0001 
include contributions from TDE debris near the central BH. Similar as what have been discussed in 
\citet{2019MNRAS.490L..81Z, 2021MNRAS.500L..57Z, 2022MNRAS.516L..66Z} for broad emission line regions 
(BLRs) if tightly associated with TDEs debris, non-virial dynamic properties of broad emission line clouds related to TDE 
debris could be expected, such as the results in the TDE candidate ASASSN-14li in \citet{2016MNRAS.455.2918H}: stronger 
emissions leading to wider line widths of broad  $H_\alpha$ which are against the results by the virialization assumptions 
to BLRs clouds. The non-virial dynamic properties of broad line clouds nearer to central BH in SDSS J0001 could be applied 
to explain the TDE model determined BH mass smaller than the virial BH mass.

In addition to the virial black hole mass and the black hole mass determined by the MOSFIT model, we also tried to estimate the black hole mass using $TDEmass$ \citep{2020ApJ...904...73R}, which calculates the masses of the black hole and the disrupted star based on the peak luminosity and temperature of the flare. Using input parameters obtained from a Gaussian rise and an exponential decay model. We found that $TDEmass$ was unable to determine the black hole mass and the disrupted star mass for SDSS J0001, as the peak luminosity and temperature fall outside the limits explored by the $TDEmass$, similar to the four TDE candidates with high luminosities above \(10^{45}\text{erg/s}\) reported by \citet{2023ApJ...942....9H}.

\subsection{Probability of the long-term variability of SDSS J0001 as intrinsic AGN variability}

Moreover, As we all know, SMBHs are common in broad line AGN, and star formations could be commonly expected around accretion flows as well discussed in \citet{2020MNRAS.493.3732D,2009Sci...323..754K}. Therefore, TDE candidates could be expected in broad-line AGN. The results mentioned above are derived from an assumption that the long-term variability of SDSS J0001 are related to a central TDE. Thus, it is necessary to evaluate the possibility that the long-term variability of SDSS J0001 results from intrinsic AGN activity. The intrinsic AGN flares can be described by the damped random walk (DRW) process (also known as the 
Ornstein–Uhlenbeck process or the Gaussian first-order continuous autoregressive (CAR) process (e.g. \citealt{2009ApJ...698..895K,2013ApJ...765..106Z,2010ApJ...708..927K, 2016ApJ...819..122Z}). Many studies have investigated AGN variability properties through 
the DRW/CAR process, such as the results well discussed in \citet{2010ApJ...721.1014M, 2012ApJ...753..106M, 2013A&A...554A.137A, 2021ApJ...922..248X, 2022MNRAS.512.5580S}. The DRW/CAR process (or stochastic processes) with damping 
timescale $\tau$ (a timescale for the time series to become uncorrelated) and intrinsic variability 
amplitude $\sigma$ ($SF_\infty\sim\sigma\sqrt{\tau}$ as the parameter used in \citet{2010ApJ...721.1014M}). To describe the 
stochastic variability of AGN light curves, the public code JAVELIN\footnote{https://github.com/nye17/javelin\#citation} has been widely 
applied.

\begin{figure*}
\centering
\includegraphics[width = 8cm,height=5cm]{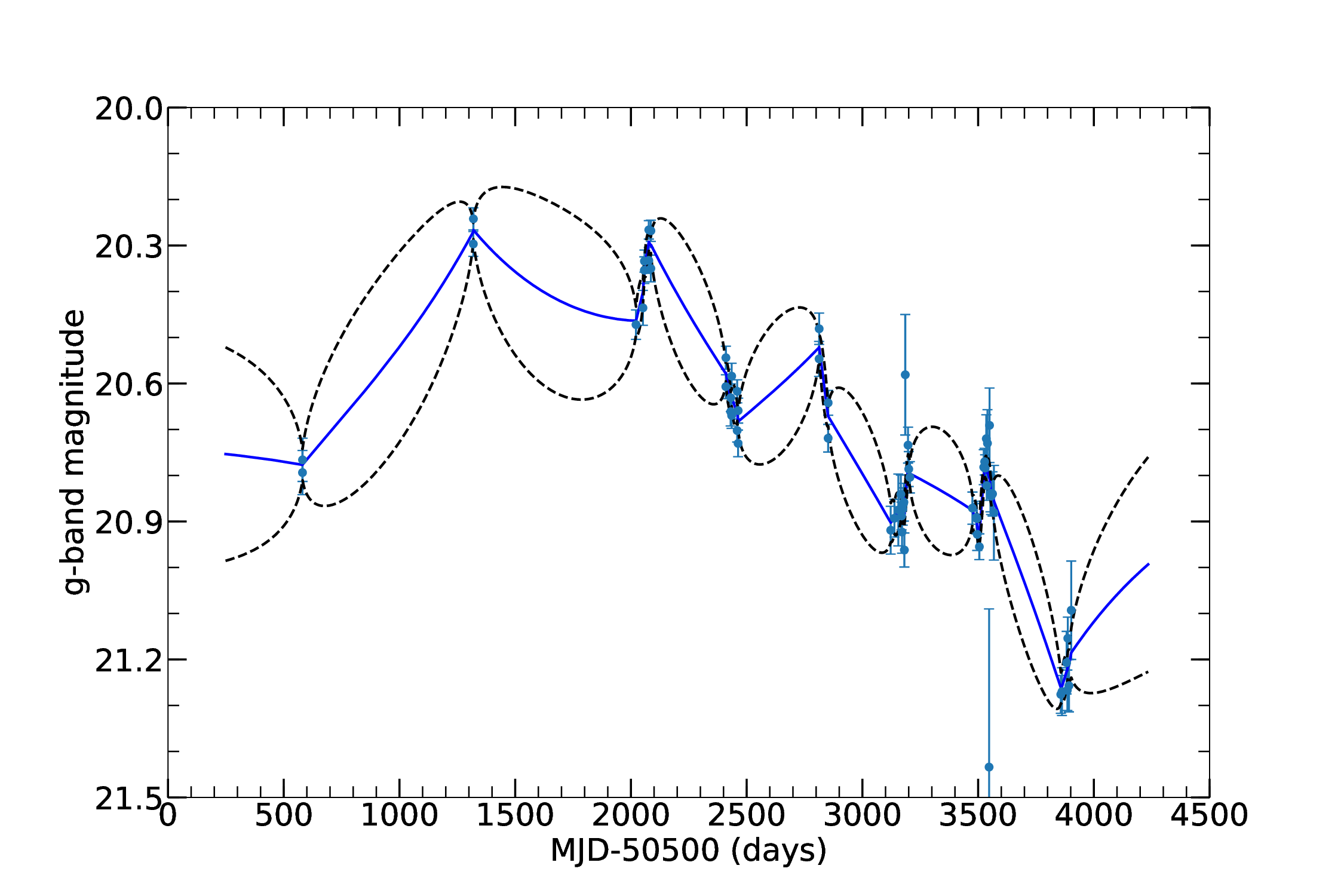}
\includegraphics[width = 8cm,height=5cm]{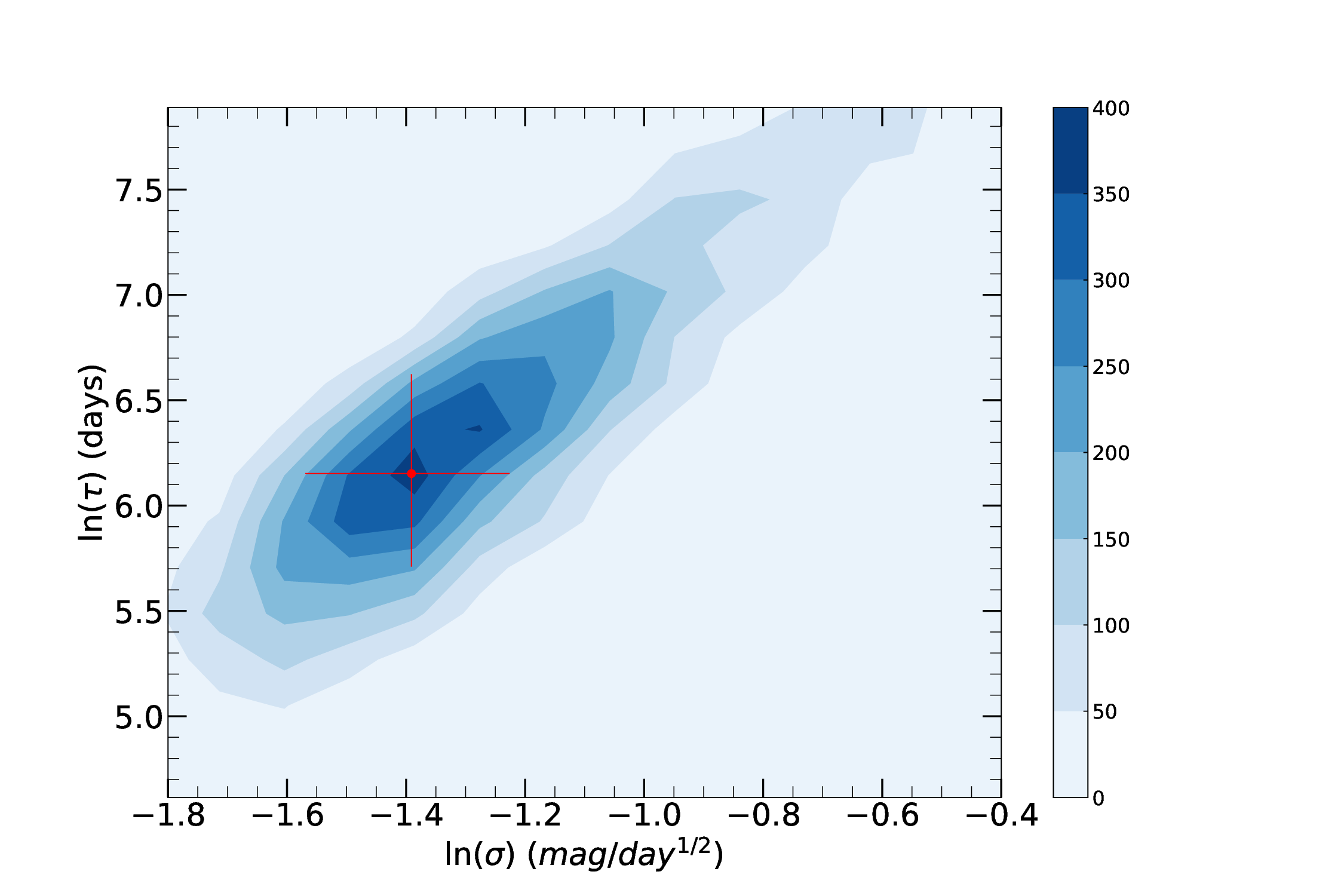}
\caption
{{\em Left panel---}The best fit with the JAVELIN code (solid blue line) and the corresponding $1\sigma$ confidence band (the black dashed lines) to the $g$-band light curve of SDSS J0001  in observer frame.
{\em Right panel---} Two-dimensional posterior distributions in the $\tau-\sigma$ plane derived from the MCMC fit, where the red circle with error bar indicate the central values and $1\sigma$ 
uncertainties of ln($\tau$) and ln($\sigma$).}
\label{7}
\end{figure*}

Then, based on the public JAVELIN code, the left panel of Figure \ref{7} shows the best fit and the corresponding 1$\sigma$ confidence band to the photometric SDSS $g$-band light curve with the JAVELIN code. The reduced $\chi^2$ of the fit is $\chi^2/dof\sim0.98$. The right panel of Figure \ref{7} shows the posterior distributions of the DRW process parameters determined through the MCMC fit. We have $\ln(\tau/{\rm days})\sim 6.15_{~-0.48}^{~+0.44}$ 
($\tau\sim470_{~-178}^{~+260}$ days) and $\ln(\sigma/({\rm mag/days}^{0.5}))\sim -1.38_{~-0.18}^{~+0.17}$. 

Before proceeding further, it is necessary to test if the DRW/CAR model
could be applied to the light curve length of the SDSS J0001. Following \citet{2010ApJ...721.1014M}, we calculate the $\Delta{L_\infty}\equiv ln({L_{best}}/{L_\infty})$, where $L_{best}$ is the likelihood of the stochastic model, $L_\infty$ is the 
likelihood that $\tau\rightarrow\infty$, indicating that the light curve length is too short to accurately measure $\tau$. We calculate the 
 $\Delta{L_\infty}=0.99$ for the  $g$-band light curve of SDSS J0001. This means that the time duration of the light curve is long enough (\citet{2010ApJ...721.1014M} exclude cases where $\Delta{L_\infty} \leq 0.05$) for accurately measuring the process parameters. On the other hand, based on the discussions in \citet{2017A&A...597A.128K} regarding the input DRW timescale ($\tau_{inp}$)
and experiment length($t_{exp}$), we calculate $\rho_{inp} \sim 0.1$ ( here, $\rho_{inp} = \tau_{inp} \times t_{exp}^{-1}$ ), 
indicating that intrinsic process 
parameters can be recovered through applications of CAR/DRW process to describe the long-term variability of SDSS J0001.

	Based on the determined process parameter $\tau$, through the CAR process discussed in \citet{2009ApJ...698..895K}, 
mock light curves $X(t)$ to trace intrinsic AGN activity can be created by 
\begin{equation}
dX(t)=- \frac{1}{\tau}X(t)+ \sigma_{*}\sqrt{dt} \epsilon (t)+bdt,
\end{equation}
where $\epsilon(t)$ represents a white noise process with zero mean and a variance of 1 (created by \emph{randomn} 
function in IDL in this analysis), \emph{bdt} is the mean magnitude of the light curve. For a given light curve, the relation between CAR process parameter $\tau$, $\sigma^{*}$, and variance of the light curve ($v$) can be estimated as $v\sim\tau\sigma_{*}^2/2$. Here, the used parameter $\sigma_{*}$ in the CAR process above has similar physical meanings as $\sigma$ applied in JAVELIN code, but $\sigma_{*}$ and $\sigma$ are not exactly equal.

\begin{figure*}
\centering
\includegraphics[width = 8cm,height=5cm]{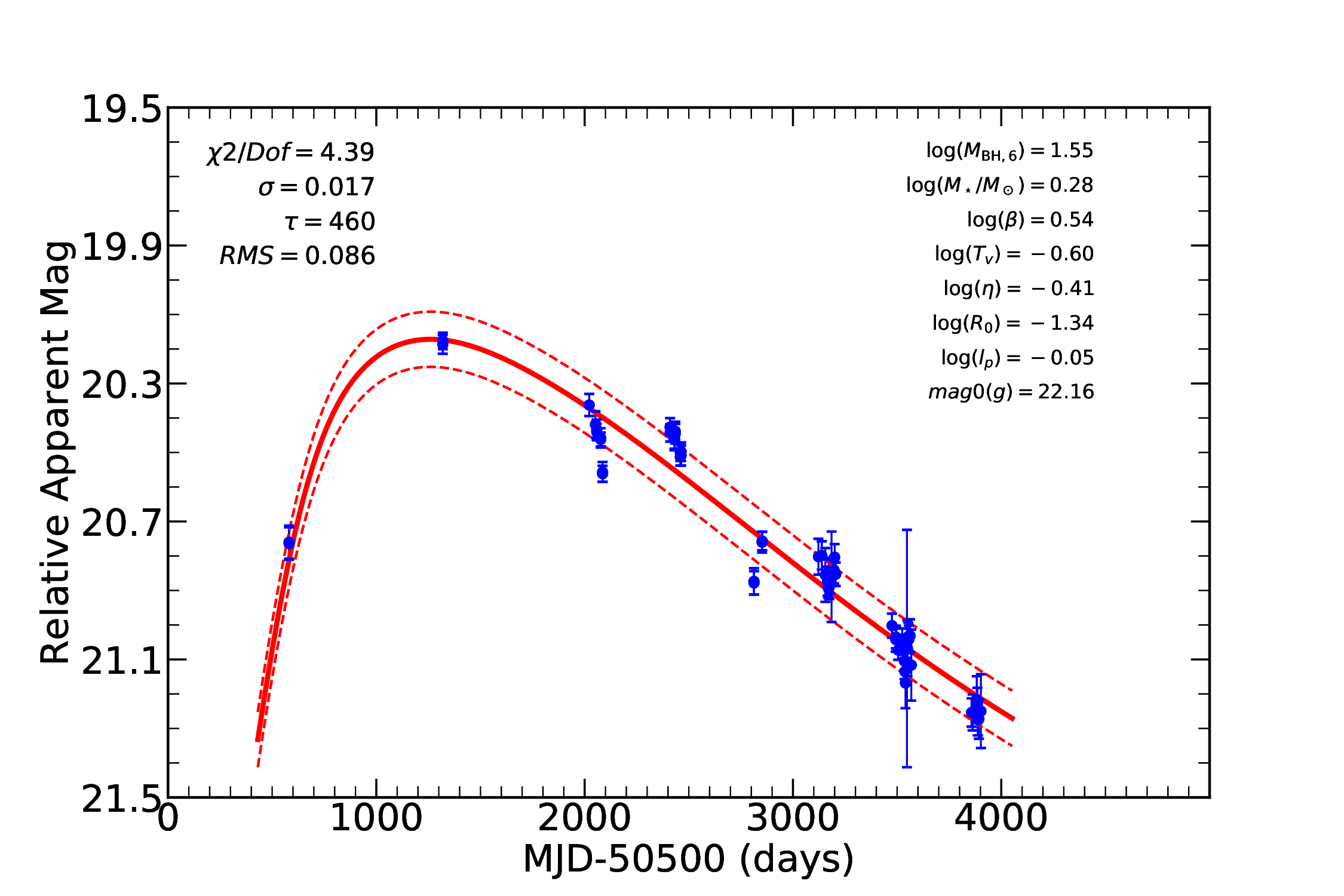}
\includegraphics[width = 8cm,height=5cm]{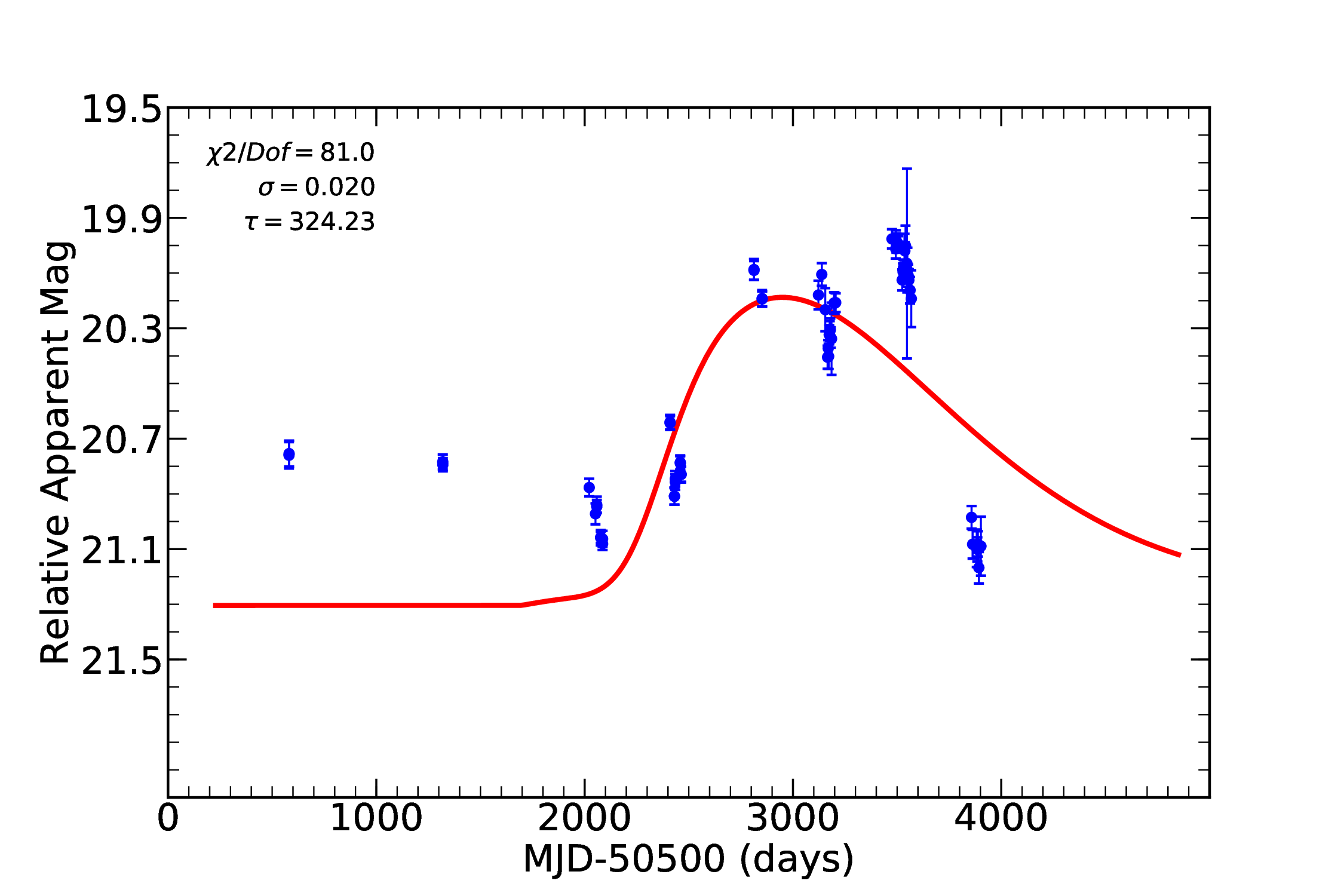}
\caption{Examples of mock light curves generated by the CAR process model that pass ({\rm left panel}) and do not pass ({\rm right panel}) the criterion for evaluating whether the light curves can be fitted with the TDE model, i.e. $\chi^2/{\rm dof}<4.5$. The red solid and dashed lines represent the best TDE model fit and the corresponding RMS as determined by the kmpfit by the kmpfit. The parameters of the CAR process model for generating the mock light curves and the parameter of the TDE model are marked in each panel.}

\label{CAR}
\end{figure*}

We use the CAR process model to generate mock $g$-band light curves by adopting $bdt=20.76$ (the mean magnitude of $g$-band light curve of SDSS J0001) and $v=0.07$ (the variance of $g$-band light curve of SDSS J0001). The CAR process parameter $\tau$ is randomly selected from 470-178 to 470+260 (determined $\tau$ plus/minus uncertainties for SDSS J0001) and the parameter $\sigma_*$ is determined by $v\sim0.07$. The mock light curves are sampled the same as the data points. The uncertainties of $F_{\rm sim }(t)$ of the mock data are simply determined as the relative error of the observational data, 
\begin{equation}
\delta F_{\rm sim }(t)=F_{\rm sim }(t) \times \frac{\delta F_{\rm obs}(t)}{F_{\rm obs}(t)},
\end{equation}
where $F_{\rm obs}$ and  $\delta F_{\rm obs}$ as the observational flux of $g$-band light curve and the corresponding uncertainty 
of SDSS J0001. Based on Equations (10) and (11), we generated $10^5$ mock light curves and fitted them with the TDE model. Since the best fit with the TDE model to the observational data of SDSS J0001 is 4.5, we selected a TDE candidate with the criterion of $\chi^2/{\rm dof}<4.5 $. This means that the TDE model fit to the selected mock light curve is comparable to that for SDSS J0001. Finally, we
found 9 light curves that can be well described by the theoretical TDE
model. Therefore, the probability is about 0.009\%. Figure \ref{CAR} displays mock light curves that pass and do not pass the criterion for evaluating whether the light curves can fit with the TDE model.

Note that simulations by \cite{2022MNRAS.516L..66Z} were made by setting the uniform distribution of $\tau\in \{50,5000\}$ days (similar to reported values of quasars in \cite{2009ApJ...698..895K,2010ApJ...721.1014M}) but calculating the $\sigma_{*}$ using the variance of the light curve of SDSS J0141. The logic of this approach is to estimate the probability that a light curve with $\tau\in \{50,5000\}$ days for normal quasars can be modeled with the TDE model. Following this approach, we calculate the upper limit of the probability $p =0.144\%$ for SDSS J0001 by adopting $\tau\in \{50,5000\}$ days in a normal distribution and using $\tau\sigma_*^2/2\sim 0.07$. In this work, we estimate the probability based on the DRW model-determined parameters and their uncertainties($\tau\sim470_{~-178}^{~+260}$) and then calculate the probability. The logic of this approach is that, if the observational data are from the variability of SDSS J0001, what is the chance probability of the data being fitted with the TDE model by considering the observational uncertainty? The derived probability is $0.009\%$.

Based on the discussions above, there are further clues to support the potential central TDE in SDSS J0001. Moreover, \citet{2023arXiv231007095X} proposed a method to search for probable hidden TDEs in normal broad line AGN with apparent intrinsic variability by considering the effects of contributions of TDEs expected variability to normal AGN variability. TDE candidates could be widely expected in broad-line AGN even with stronger activities. In other words, before detecting 
and reporting hidden TDEs in normal broad-line AGN with apparent variability, preliminary clues can be provided to study the 
connections between intrinsic AGN activity and variability related to TDEs. This can be achieved by detecting and reporting more 
TDE candidates in broad-line AGN (quasars) without apparent long-term variability outside of the time durations of expected TDEs. 
This is our main objective in the current stage.

\subsection{\large Alternative explanations for the flare observed in SDSS J0001}
Other possibilities may also explain the flare shown in the first five panels of Figure~\ref{1} as discussed by 
\citep{2016MNRAS.457..389M,2016MNRAS.463..296L,2017MNRAS.470.4112G, 2016MNRAS.457..389M}. 

\begin{figure*}
\centering

\includegraphics[width = 7cm,height=5cm]{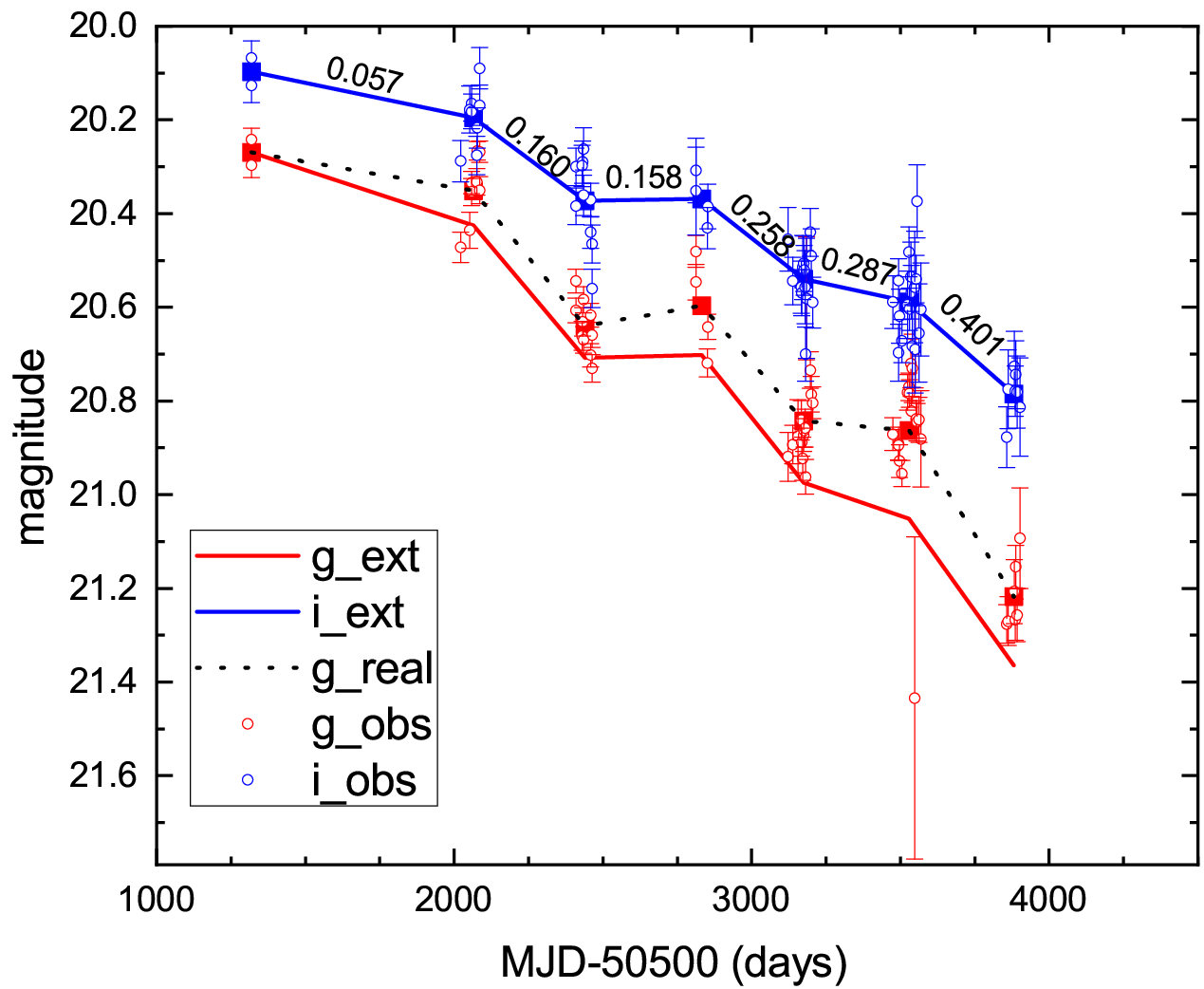}
\hspace{0.6cm}
\includegraphics[width = 7cm,height=5cm]{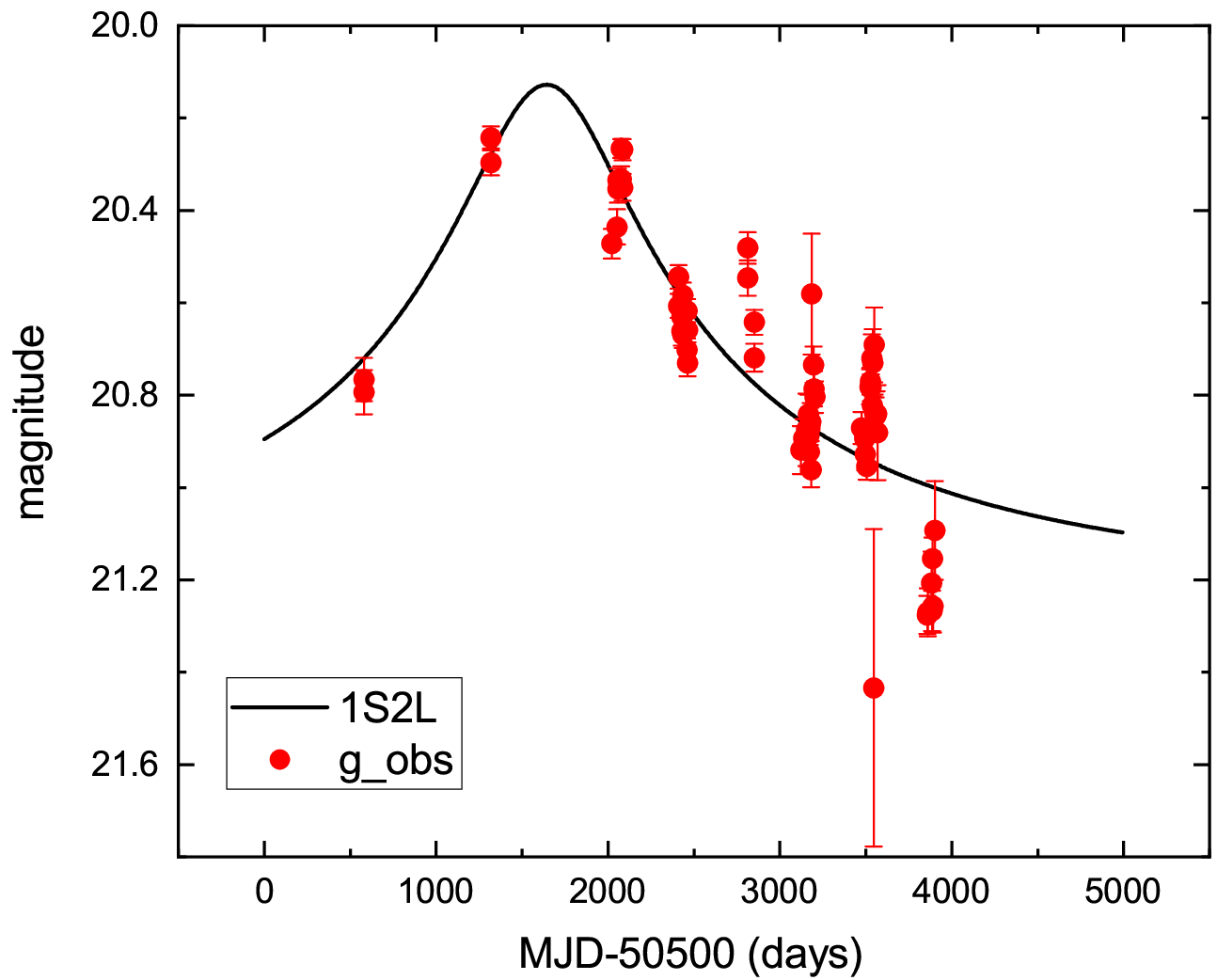}
\caption
{{\em Left panel---} The dust extinction effect on the $g$-band (red open circles) and $i$-band (blue open circles) data of SDSS J0001. 
The solid blue (or red) squares represent the average values of the $i$-band ($g$-band) magnitude  in the following bins with zero time at MJD-50500: [1000, 1500], [1500, 2200], [2200, 2600], [2600, 3000], [3000,3300], [3300,3700], and [3700, 4000]. The blue solid squares from the $i$-band light curve are applied to determine
\emph{E(B-V)} values between two epochs, marked as black characters above the solid blue line which is applied to linearly connect the blue solid squares. The solid red line shows the expected light curve through the excitation effects with applications of the same \emph{E(B-V)} values from the $i$-band light curve. The right panel shows the best-fitting results (black solid line) to the $g$-band light curve with the 1S2L model.
}
\label{9}
\end{figure*}

One possible explanation is a change of dust extinction, such as obscuration effect 
by moving dust clouds as discussed by \citet{2015ApJ...800..144L} for the TDE candidate of SDSS J0159. Here, the effects of change in dust extinction 
in SDSS J0001 are simply discussed as follows through the python package of 
dust\_extinction \footnote{https://dust-extinction.readthedocs.io/en/stable/index.html}. 
First, we assume that there is no dust extinction at the peak luminosity and that varying dust extinctions lead to the 
subsequent variability of light curve. Based on the SDSS $i$-band light curve, we can obtain the \emph{E(B-V)} values of 
dust extinctions between two near epochs, as shown by the blue curve in left panel of Figure \ref{9}. 
If accepted change in dust extinction to explain the flare shown in left panel of Figure \ref{9} in SDSS J0001, the \emph{E(B-V)} values 
could be totally same in the other bands. However, as shown in left panel in Figure \ref{9}, the same \emph{E(B-V)} values 
applied in SDSS $g$-band (red solid line) can lead to different variability from the observational results (black dot line). Specifically, the observed light 
curve is substantially higher than predicted in a scenario where dust extinction alone is responsible 
for the light curve state change, which implies that the observed photometric variability is not only caused by change in dust 
extinction. Throughout this process, the dust extinction curve of F99 is adopted \citep{1999PASP..111...63F}. Meanwhile, we 
have checked other extinction curves listed in the dust\_extinction package
and obtained similar results.

Another possibility for explaining the light curves of SDSS J0001 is the microlensing by one or multiple foreground 
stars. We used the open-source microlens light curve analysis tool, {\em MulensModel}, to describe the 
SDSS g-band light curve of SDSS J0001 accepted one point source binary lens model (1S2L) in MulensModel. The fitting results 
are shown in the right panel of Figure \ref{9}. However, considering totally similar magnification factors in different optical bands in microlensing model, there 
should be very similar variability profiles in different bands, which is against the results shown in the first five panels of Figure~\ref{1} with much 
wider variability profile in the $i$-band than the other SDSS bands. Therefore, the microlensing model should be not totally preferred in 
SDSS J0001, unless there were very distinct structures of emission regions for the SDSS $ugriz$-band emissions in SDSS J0001.

In addition, accretion can be considered to explain the light curves of SDSS J0001. On the one hand, several possible accretion models were discussed in \cite{2016MNRAS.457..389M}, but most predict longer timescales ($\sim10^4$ years), which do not match the timescale observed for SDSS J0001. It cannot be ruled out that some kind of rare eruptive accretion could explain the variability of SDSS J0001, but it is hard to judge without more detailed models. Meanwhile, many studies have shown that AGN variability can be simulated using the DRW/CAR process and is considered to represent that produced by accretion. In this manuscript, we find only a 0.009\% probability that the long-term variability in SDSS J0001 is due to central AGN accretion.

\section{Summary and CONCLUSIONS}
We have presented observational and theoretical analyses with the TDE model to a high-$z$ ($z=1.404$) TDE candidate in SDSS J0001, which is a quasar with apparent broad Mg~{\sc ii} emission line. We obtained its nine-year (MJD 51082 to 54402) photometric $ugriz$-band light curves from the SDSS
Stripe82 and the PHOTOOBJALL databases. It is found that the long-term variability of SDSS J0001 can be described by the conventional TDE model. We summarize our analyses as follows. 
\begin{itemize}
\item
The long-term variability of SDSS J0001 illustrates as a clear rise-to-peak followed by a smooth decline. The light curves can be fitted with the phenomenological TDE models. 
\item The $ugriz$-band light curves of SDSS J0001 can be described by the conventional TDE model with a main-sequence star 
of $M_\star\sim1.905_{-0.009}^{+0.023} {\rm M_\odot}$ tidally disrupted by central BH with $M_{\rm BH}\sim6.5_{-2.6}^{+3.5} \times10^7{\rm M_\odot}$. Our analysis shows that the extremely long variability timescale of the SDSS J0001 is due to its high impact parameter, large masses of the central BH and the large stellar mass, and the time-dilation effect due to its high-$z$ nature.       
\item Since the long-term variability has been detected in a quasar, we examine whether the variability results from AGN activity. Through the CAR process applied to create $10^5$ mock light curves
to trace intrinsic AGN activities, the probability of such a long-term variability being from a central AGN activity is $0.009\%$. Alternative explanations for the long-term variability, including dust extinction and microlensing, are also discussed.   

\item  The estimated virial BH mass through the broad Mg~{\sc ii} emission line in SDSS J0001 is 7.5 times larger than that derived from our analysis. It is possible that the TDE fallback accreting debris makes a significant contribution to the Mg~{\sc ii} emission clouds.

\end{itemize}
Based on our analysis, we suggest that there is a high-redshift TDE candidate in the quasar SDSS J0001. This provides a clue that TDEs can be detected in broad-line AGNs as well as in quiescent galaxies.

\section*{Acknowledgements}
We gratefully acknowledge the anonymous referee for giving us constructive comments and suggestions to greatly
improve our paper.
We sincerely thank the developer of MulensModel, 
Dr. Radek Poleski, for friendly discussions on running this package in the Win platform.
We thank Xiao Li, Xinzhe Wang, Qi Zheng, and Xiaoyan Li for useful discussion.
This work is supported by the National Natural Science Foundation of China (grants NSFC-12173020, 12373014 and 
12133003). Gu gratefully thank the kind financial support from the Innovation Project of Guangxi Graduate Education. The paper has made use of 
the code of TDEFIT \url{https://tde.space/tdefit/} which is a piece of 
open-source software written by James Guillochon for the purposes of model-fitting photometric light curves 
of tidal disruption events and also made use of the code of MOSFIT (Modular Open Source Fitter for Transients) 
\url{https://mosfit.readthedocs.io/} which is a Python 2.7/3.x package for fitting, sharing, and estimating 
the parameters of transients via user-contributed transient models. The paper has made use of the MCMC code 
\url{https://emcee.readthedocs.io/en/stable/index.html}, and made use of the Kmpfit module in Python package Kapteyn 
\url{https://www.astro.rug.nl/software/kapteyn/kmpfittutorial.html}.

%%%%%%%%%%%%%%%%%%%%%%%%%%%%%%%%%%%%%%%%%%%%%%%%%%
\section*{Data Availability}
The data underlying this article will be shared on reasonable request to the corresponding author
(\href{mailto:xgzhang@gxu.edu.cn}{xgzhang@gxu.edu.cn}).

\bibliography{reference} 
\bibliographystyle{mnras}

\section{Appendix}
\subsection{Appendix A: SQL search}

According to the multi-epoch objids related to the THINGID=107540835 for SDSS J0001, the detailed query in 
SQL \footnote{
\url{https://skyserver.sdss.org/dr16/en/tools/search/sql.aspx}}search is as follows:
\begin{sloppypar}
\noindent\rule{85mm}{0.5mm}
\\
\noindent \textbf{select} mjd, psfmag\_u, psfmagerr\_u, psfmag\_g, psfmagerr\_g, psfmag\_r, psfmagerr\_r, 
psfmag\_i,\\ psfmagerr\_i, psfmag\_z, psfmagerr\_z\\
\noindent \textbf{from} PHOTOOBJALL\\
\noindent \textbf
{where} \\
objid = 1237663277927891217
or objid=1237663479807475923\\ or objid=1237663527055327473
or objid=1237663716014489812\\ or objid=1237663784734425357or objid=1237666649495699593\\
or objid=1237667173463163111 or objid=1237646012157001880\\ or objid=1237653012973486327
or objid=1237659756054577365\\ or objid=1237659906395013297 or objid=1237660026651410679\\
or objid=1237660224222068977 or objid=1237660357358125285\\
\noindent\rule{85mm}{0.5mm}
\end{sloppypar}

\subsection{Appendix B: An example linear interpolation processes}
\setcounter{figure}{0}
\renewcommand{\thefigure}{B\arabic{figure}}
Figure \ref{int} shows examples of our calculations for $\dot{M}_{a}(T_{v}, \beta)$ by adopting some parameter 
sets of $\{T_{v}, \beta\}$. 
\begin{figure*}
\centering
\includegraphics[angle=0,scale=0.15]{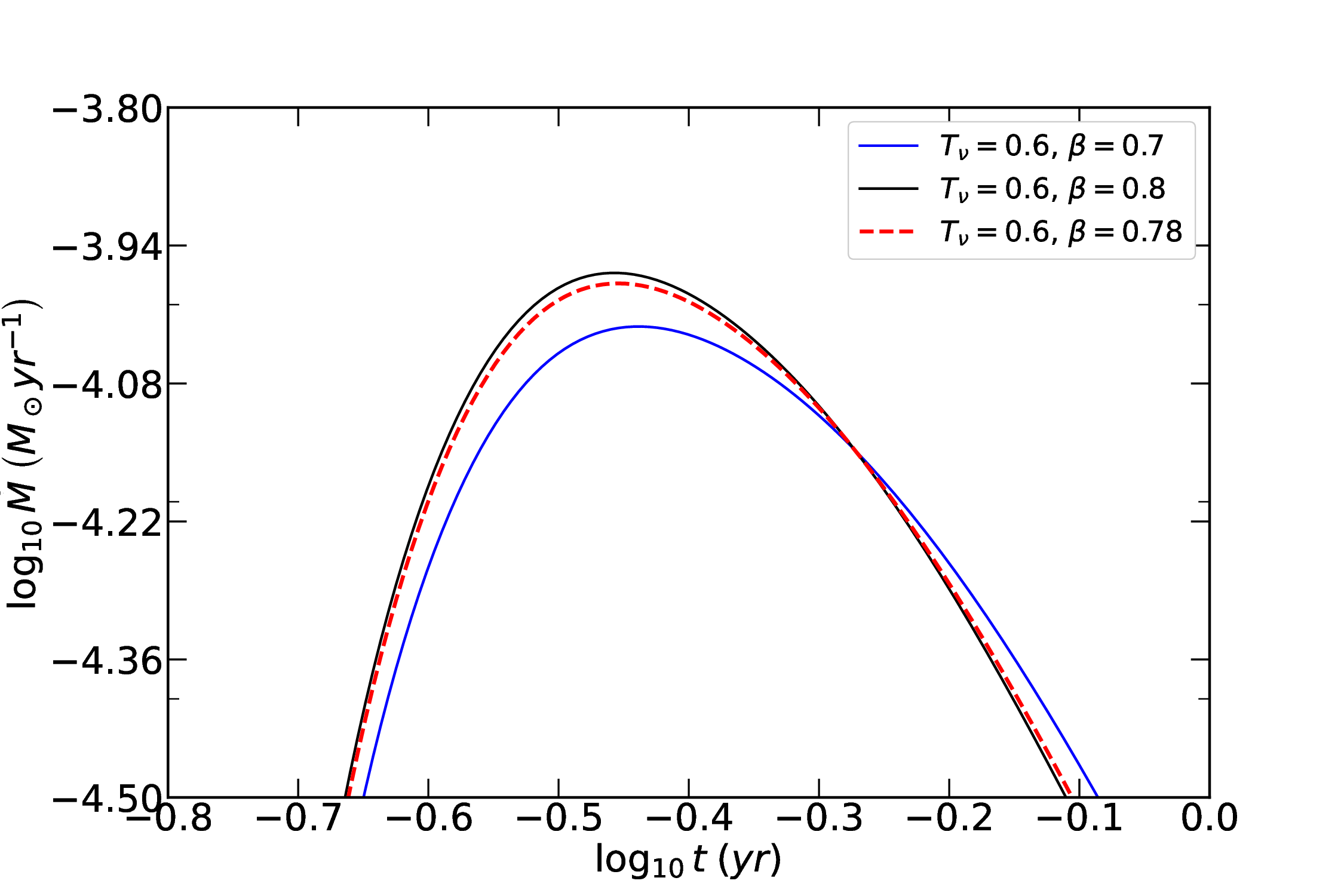}
\includegraphics[angle=0,scale=0.15]{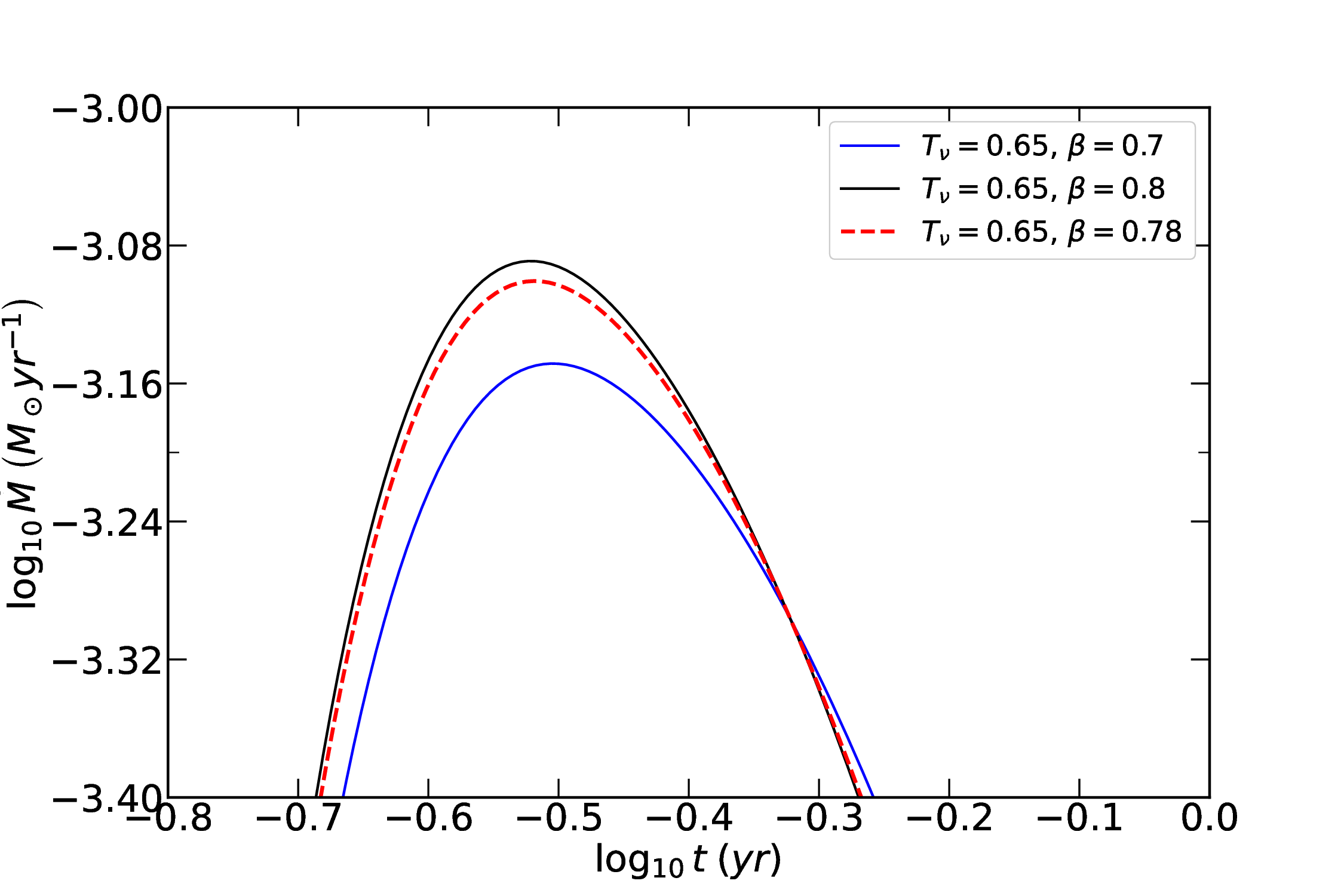}
\includegraphics[angle=0,scale=0.15]{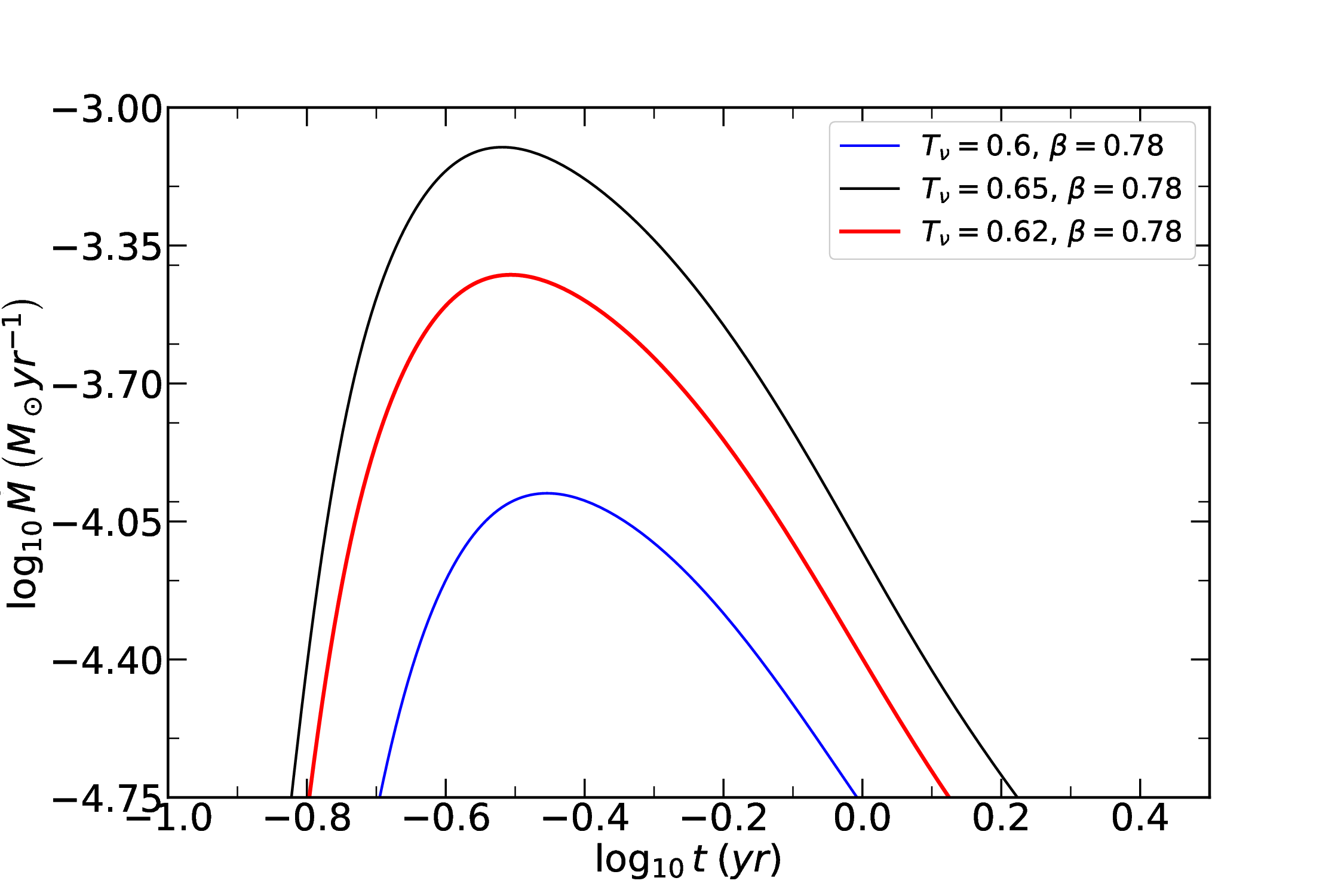}
\caption
{The intuitive schematic diagram of the interpolation process. In the left panel, the solid 
green lines represent results for $\dot{M}_{a}(T_{v}=0.6, \beta=0.7)$ and $\dot{M}_{a}(T_{v}=0.6, \beta=0.8)$ 
included in the template, and the dashed red line shows the interpolation results for $\dot{M}_{a}(T_{v}=0.6, 
\beta=0.78)$ derived from the first sub-equation in Equation (3). In the middle panel, the solid green 
lines show the results for $\dot{M}_{a}(T_{v}=0.65, \beta=0.7)$ and $\dot{M}_{a}(T_{v}=0.65, \beta=0.8)$ 
included in the template, and the dashed red line shows the interpolation results for $\dot{M}_{a}(T_{v}=0.65, 
\beta=0.78)$ from the second sub-equation in Equation (3). In the right panel, the solid green 
lines show the interpolation results for $\dot{M}_{a}(T_{v}=0.6, \beta=0.78)$ and $\dot{M}_{a}(T_{v}=0.65, 
\beta=0.78)$ shown as dashed red lines in the left and middle panels. The solid red line shows the 
interpolation results for $\dot{M}_{a}(T_{v}=0.62, \beta=0.78)$ by the Equation (4). }
\label{int}
\end{figure*}

\bsp	
\label{lastpage}
\end{document}